\documentclass[twocolumn,showpacs,preprintnumbers,amsmath,amssymb,pra,superscriptaddress]{revtex4}
\usepackage[latin1]{inputenc}
\usepackage{dcolumn,color}
\usepackage{bm}
\usepackage{graphicx}
\usepackage[english]{babel}
\usepackage[bbgreekl]{mathbbol}
\usepackage{bbm}
\usepackage{stmaryrd}

\DeclareMathOperator{\Tr}{Tr}
\DeclareMathOperator{\Rea}{Re}

\DeclareMathOperator{\diag}{diag}
\def\bbm[#1]{\mbox{\boldmath $#1$}}
\newcommand{\ket}[1]{\displaystyle{|#1\rangle}}
\newcommand{\bra}[1]{\displaystyle{\langle #1|}}
\newcommand{\TE}{\text{TE}}
\newcommand{\TM}{\text{TM}}

\makeatletter
\def\amsbb{\use@mathgroup \M@U \symAMSb} %definisco amsbb anziché mathbb per il blackboard bold di ams
\makeatother

\usepackage{hyperref}
\hypersetup{colorlinks,
linkcolor=black,
filecolor=green,
urlcolor=blue,
citecolor=black}

\begin{document}

\title{Casimir-Lifshitz force out of thermal equilibrium between dielectric gratings}

\author{Antonio Noto}
\affiliation{Universit\'{e} Montpellier 2, Laboratoire Charles Coulomb UMR 5221 - F-34095, Montpellier, France}\affiliation{CNRS, Laboratoire Charles Coulomb UMR 5221 - F-34095, Montpellier, France}
\affiliation{Dipartimento di Fisica e Chimica dell'Universit\`{a} degli Studi di Palermo and CNISM, Via Archirafi 36, I-90123 Palermo, Italy}
\author{Riccardo Messina}
\author{Brahim Guizal}
\affiliation{Universit\'{e} Montpellier 2, Laboratoire Charles Coulomb UMR 5221 - F-34095, Montpellier, France}\affiliation{CNRS, Laboratoire Charles Coulomb UMR 5221 - F-34095, Montpellier, France}
\author{Mauro Antezza}
\affiliation{Universit\'{e} Montpellier 2, Laboratoire Charles Coulomb UMR 5221 - F-34095, Montpellier, France}\affiliation{CNRS, Laboratoire Charles Coulomb UMR 5221 - F-34095, Montpellier, France}
\affiliation{Institut Universitaire de France - 103, bd Saint-Michel - F-75005 Paris, France}

\date{\today}

\begin{abstract}
We calculate the Casimir-Lifshitz pressure in a system consisting of two different 1D dielectric lamellar gratings having two different temperatures and immersed in an environment having a third temperature. The calculation of the pressure is based on the knowledge of the scattering operators, deduced using the Fourier Modal Method. The behavior of the pressure is characterized in detail as a function of the three temperatures of the system as well as the geometrical parameters of the two gratings. We show that the interplay between non-equilibrium effects and geometrical periodicity offers a rich scenario for the manipulation of the force. In particular, we find regimes where the force can be strongly reduced for large ranges of temperatures. Moreover, a repulsive pressure can be obtained, whose features can be tuned by controlling the degrees of freedom of the system. Remarkably, the transition distance between attraction and repulsion can be decreased with respect to the case of two slabs, implying an experimental interest for the observation of repulsion.
\end{abstract}

\pacs{12.20.-m, 42.79.Dj, 42.50.Ct, 42.50.Lc}

\maketitle

\section{Introduction}

Casimir-Lifshitz force in an interaction originating from the fluctuations of the electromagnetic field and existing between any couple of polarizable bodies. It was first theoretically derived by Casimir in 1948 \cite{CasimirProcKNedAkadWet48,CasimirPhysRev48} in the idealized configuration of two perfectly conducting parallel plates at zero temperature. Later, Lifshitz and collaborators generalized the calculation to the case of bodies having arbitrary optical properties and of finite temperature \cite{DzyaloshinskiiAdvPhys61}. The Casimir-Lifshitz interaction, experimentally verified for several different geometries \cite{CasimirDalvit}, results from two contributions, one originating from vacuum fluctuations and present also at zero temperature, the other one from purely thermal fluctuations. The latter becomes relevant when the distance separating the bodies is larger than the thermal wavelength $\lambda_T=\hbar c/k_\text{B}T$, of the order of $8\,\mu$m at ambient temperature. This explains why it has been only very recently experimentally observed at thermal equilibrium \cite{SushkovNatPhys11}.

Nevertheless, the situation completely changes out of thermal equilibrium. It was first theoretically predicted in 2005 that the atom-surface interaction (usually referred to as Casimir-Polder force) is qualitatively and quantitatively modified with respect to thermal equilibrium \cite{AntezzaPRL05,AntezzaJPhysA06}. New power-law behaviors appear, the force can turn into repulsive (being only attractive at thermal equilibrium) and it is strongly tunable by modifying the temperatures involved in the system. This prediction was verified in 2007, providing the first experimental observation of thermal effects \cite{ObrechtPRL07}. These results paved the way to a renewed interest in Casimir-Lifshitz effects out of thermal equilibrium. In fact, this effect was studied for two slabs \cite{AntezzaPRL06,AntezzaPRA08} and in presence of atoms \cite{AntezzaPRA04,BuhmannPRL08,SherkunovPRA09,BehuninPRA10,BehuninJPhysA10,BehuninPRA11}, and more recently several different approaches have been developed to deal with the problem of the force out of thermal equilibrium and heat transfer between two \cite{BimontePRA09,MessinaEurophysLett11,MessinaPRA11,KrugerEurophysLett11,KrugerPRB12,RodriguezPRL11,McCauleyPRB12,RodriguezPRB12} or more \cite{KrugerPRL11,RodriguezPRB13,MessinaPRA14} arbitrary bodies. The physics of the electromagnetic field out of thermal equilibrium has also stimulated the study of other effects, such as the manipulation of atomic populations \cite{BellomoEurophysLett12,BellomoPRA13} and entanglement \cite{BellomoEurophysLett13,BellomoNewJPhys13}.

\begin{center}\begin{figure}[htb]
\includegraphics[width=8cm]{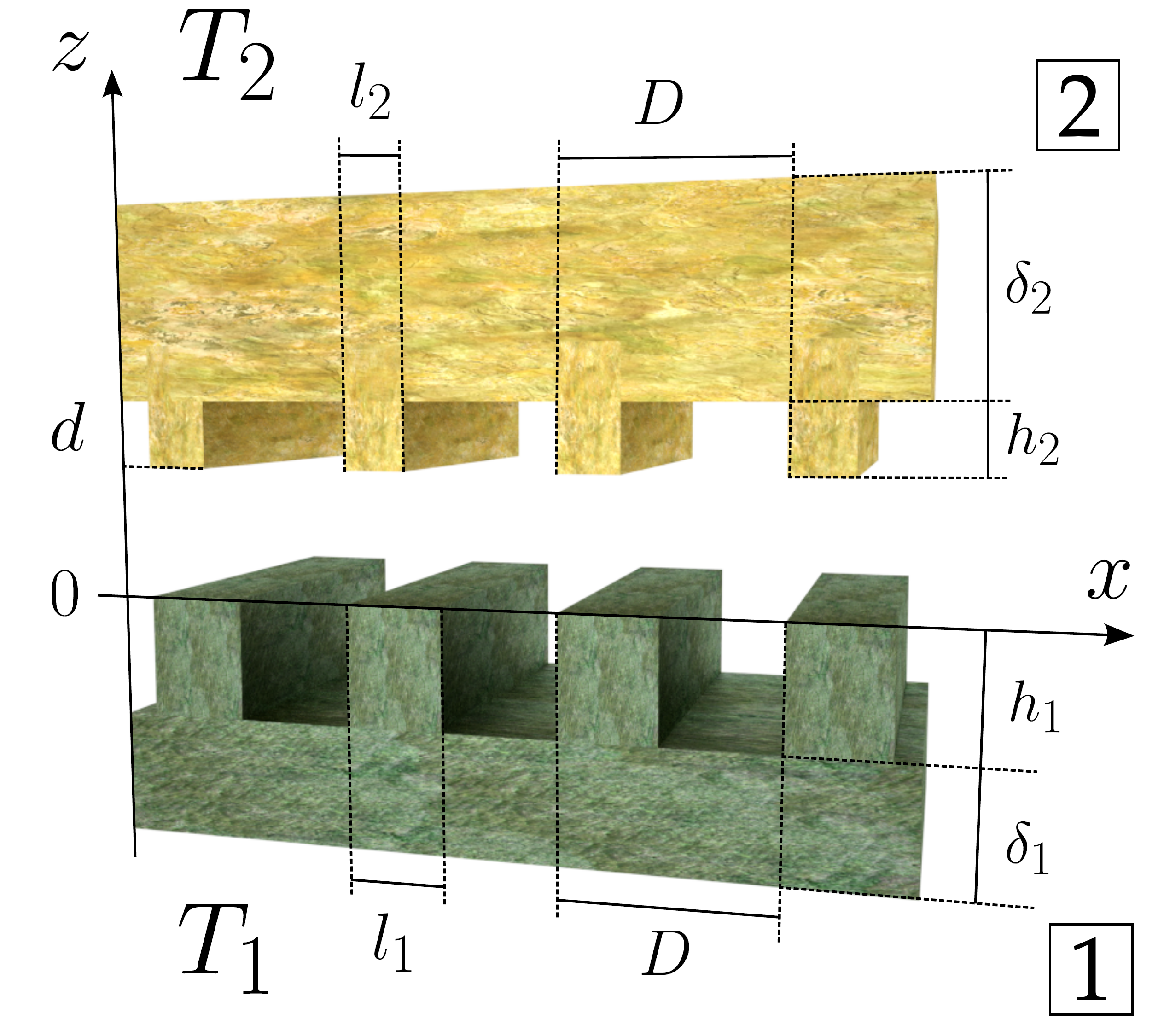}
\caption{(Color online) Geometry of the system. Two gratings, labeled with 1 and 2, at a distance $d$, always assumed to be positive. The gratings, in general made of different materials, are infinite in the $xy$ plane, and periodic in the $x$ direction with the same period $D$. They have corrugation depths $h_i$ ($i=1,2$), thicknesses $\delta_i$ and lengths of the upper part of the grating $l_i$. This defines the filling factors $f_i=l_i/D$.}\label{Fig1}\end{figure}\end{center}

In parallel with the interest in the absence of thermal equilibrium, Casimir-Lifshitz interactions have been studied in several different geometries, with particular interest in the sphere-plane configuration, the most studied experimentally. More recently, nanostructured surfaces have been theoretically considered in the contexts of both force \cite{DavidsPRA10,IntravaiaPRA12,LussangePRA12,GueroutPRA13} and heat transfer \cite{LussangePRB12,GueroutPRB12}. Experimentally, the force have been measured between a sphere and a dielectric \cite{ChanPRL08,BaoPRL10} or metallic \cite{IntravaiaNatComm14} grating.

The problem we address here for the first time is the calculation of the Casimir-Lifshitz force out of thermal equilibrium in presence of dielectric gratings, in order to study the combination of non-equilibrium and geometrical effects. In particular we consider a system made of two different gratings having different temperatures, immersed in an environmental bath at a third temperature. Our calculations can be relevant both to imagine new experiments measuring the Casimir-Lifshitz force out of thermal equilibrium and in the more general context of the manipulation of the force in micro- and nano-electromechanical systems \cite{ChanScience01,ZouNatComm13}.

The paper is structured as follows. In Sec. \ref{SecIntro} we introduce our physical system and provide the notation and main definitions. In Sec. \ref{SecFMM} we solve the problem of the scattering upon a single 1D lamellar dielectric grating using the Fourier Modal Method. In Sec. \ref{SecNum}, we apply these results in order to calculate the force out of thermal equilibrium between two different gratings. We explore the behavior of the force as a function of the three temperatures and of the geometrical parameters of the gratings, with a specific attention to the appearance and features of repulsion. We finally give in Sec. \ref{SecConcl} some conclusive remarks.

\section{Physical system and force out of thermal equilibrium}\label{SecIntro}

We start by describing the system studied in this paper. We address the Casimir-Lifshitz force between two dielectric gratings immersed in vacuum ($\varepsilon=1$) in the geometrical configuration shown in Fig. \ref{Fig1}. We label the two gratings with an index $i$ taking values 1 and 2. The gratings are infinite in $x$ and $y$ directions, with periodicity along the $x$ axis. Their distance $d$ is defined in Fig. \ref{Fig1} and can only take positive values (i.e. a plane $z=\bar{z}$ must exist separating the two bodies). The gratings share the same period $D$ and have corrugation depth $h_i$, permittivities $\varepsilon_i(\omega)$ in the homogeneous zone, permittivities $\varepsilon_i(x,\omega)$ along the grating zone having thickness $\delta_i$, and filling factors $f_i=l_i/D$ ($l_i$ is defined as in Fig. \ref{Fig1}).

Our physical system is considered in a configuration out of thermal equilibrium (OTE). This means that each body is supposed to be in local thermal equilibrium with a constant temperature $T_i$. We also assume that the two gratings are immersed in a radiation bath coming from bounding walls far from the system and having temperature $T_\text{e}$, in general different from the temperatures of the two gratings (see also \cite{MessinaPRA11}). The whole system is considered in a stationary regime so that the three temperatures involved are constant in time.

In \cite{MessinaEurophysLett11,MessinaPRA11,MessinaPRA14}, this assumption has been used to characterize the properties of the source fields (the ones emitted by the two bodies and coming from the surrounding walls) in terms of field correlation functions. This procedure is based on a mode decomposition of the fields, each mode $(\omega,\mathbf{k},p,\phi)$ being identified by the direction of propagation $\phi=+,-$ along the $z$ axis, the polarization index $p$ [assuming the values $p=1,2$ which respectively correspond to transverse electric (TE) and transverse magnetic (TM) modes], the frequency $\omega$ and the transverse wavevector $\mathbf{k}=(k_x,k_y)$. In this description, the $z$ component of the wavevector $k_z$ is a dependent variable defined as
\begin{equation}\label{eq:kz}
k_{z}=\sqrt{\frac{\omega^2}{c^2}-\mathbf{k}^2}.
\end{equation}
Based on this mode decomposition, the trace of a given operator $\mathcal{O}$ is defined as
\begin{equation}\label{DefTrace}
\Tr\mathcal{O}=\sum_p\int\frac{d^2\mathbf{k}}{(2\pi)^2}\int_0^{+\infty}\frac{d\omega}{2\pi}\bra{p,\mathbf{k}}\mathcal{O}\ket{p,\mathbf{k}}.
\end{equation}
The correlation functions of the field have been expressed as a function of the reflection and transmission operators $\mathcal{R}$ and $\mathcal{T}$ associated to each body (see Sec. \ref{SecFMM} for more details). Using these correlation functions, the OTE Casimir-Lifshitz force acting on body 1 can be cast in the following form \cite{MessinaPRA11,MessinaPRA14} (the distance dependence is implicit):
\begin{equation}\label{GenForce}
F_{1z}=F_{1z}^{\text{(eq)}}(T_1)+\Delta(T_1,T_2,T_\text{e}),
\end{equation}
where $F_{1z}^{\text{(eq)}}(T_1)$ is the force acting on body 1 at thermal equilibrium at its temperature $T_1$. This equilibrium contribution reads
\begin{align}\label{Feq}
F_{1z}^{\text{(eq)}}&=-2\Rea\Tr\Bigl[k_z\omega^{-1}N(\omega,T)\nonumber\\
&\,\times\Bigl(U^{(12)}\mathcal{R}^{(1)+}\mathcal{R}^{(2)-}+U^{(21)}\mathcal{R}^{(2)-}\mathcal{R}^{(1)+}\Bigr)\Bigr],
\end{align}
while the non-equilibrium term is
\begin{widetext}
\begin{align}\label{Delta}
&\Delta(T_1,T_2,T_\text{e})=-\hbar\Tr\Bigl\{\Bigl[n_{e1}\Bigl[U^{(21)}\mathcal{T}^{(2)-}\mathcal{P}_{-1}^{\text{(pw)}}\mathcal{T}^{(2)-\dag}U^{(21)\dag}\Bigl(f_2(\mathcal{R}^{(1)+})-\mathcal{T}^{(1)-\dag}\mathcal{P}_2^{\text{(pw)}}\mathcal{T}^{(1)-}\Bigr)\nonumber\\
&+\Bigl(U^{(12)}\mathcal{T}^{(1)+}\mathcal{P}_{-1}^{\text{(pw)}}\mathcal{T}^{(1)+\dag}U^{(12)\dag}-\mathcal{P}_{-1}^{\text{(pw)}}\Bigr)f_2(\mathcal{R}^{(2)-})+\Bigl(\mathcal{R}^{(2)-}\mathcal{P}_{-1}^{\text{(pw)}}\mathcal{R}^{(2)-\dag}-\mathcal{R}^{(12)-}\mathcal{P}_{-1}^{\text{(pw)}}\mathcal{R}^{(12)-\dag}\Bigr)\mathcal{P}_2^{\text{(pw)}}\Bigr]\nonumber\\
&+n_{21}U^{(21)}\Bigl(f_{-1}(\mathcal{R}^{(2)-})-\mathcal{T}^{(2)-}\mathcal{P}_{-1}^{\text{(pw)}}\mathcal{T}^{(2)-\dag}\Bigr)U^{(21)\dag}\Bigl(f_2(\mathcal{R}^{(1)+})-\mathcal{T}^{(1)-\dag}\mathcal{P}_2^{\text{(pw)}}\mathcal{T}^{(1)-}\Bigr)\Bigr]\Bigr\}.
\end{align}
\end{widetext}
In the equations above we have introduced the thermal population
\begin{equation}N(\omega,T)=\frac{\hbar\omega}{2}\coth\Bigl(\frac{\hbar\omega}{2k_\text{B}T}\Bigr)=\hbar\omega\Bigl[\frac{1}{2}+n(\omega,T)\Bigr],\end{equation}
with
\begin{equation}n(\omega,T)=\frac{1}{e^{\frac{\hbar\omega}{k_\text{B}T}}-1},\end{equation}
and the population differences $n_{ij}=n(\omega,T_i)-n(\omega,T_j)$. Moreover we have defined the auxiliary functions
\begin{equation}
f_\alpha(\mathcal{R})=
\begin{cases}
\mathcal{P}_{-1}^{\text{(pw)}}-\mathcal{R}\mathcal{P}_{-1}^{\text{(pw)}}\mathcal{R}^\dag+\mathcal{R}\mathcal{P}_{-1}^{\text{(ew)}}-\mathcal{P}_{-1}^{\text{(ew)}}\mathcal{R}^\dag\\
 \hspace{5.3cm}\alpha=-1,\\
\mathcal{P}_2^{\text{(pw)}}+\mathcal{R}^\dag\mathcal{P}_2^{\text{(pw)}}\mathcal{R}+\mathcal{R}^\dag\mathcal{P}_2^{\text{(ew)}} +\mathcal{P}_2^{\text{(ew)}}\mathcal{R}\\
\hspace{5.3cm}\alpha=2.
\end{cases}
\end{equation}
and the operators
\begin{align}
&U^{(12)}=\sum_{n=0}^{+\infty}\bigl(\mathcal{R}^{(1)+}\mathcal{R}^{(2)-}\bigr)^n=(1-\mathcal{R}^{(1)+}\mathcal{R}^{(2)-})^{-1},\\
&U^{(21)}=\sum_{n=0}^{+\infty}\bigl(\mathcal{R}^{(2)-}\mathcal{R}^{(1)+}\bigr)^n=(1-\mathcal{R}^{(2)-}\mathcal{R}^{(1)+})^{-1},\\
&\mathcal{R}^{(12)-}=\mathcal{R}^{(1)-}+\mathcal{T}^{(1)-}U^{(21)}\mathcal{R}^{(2)-}\mathcal{T}^{(1)+}.
\end{align}
Finally, in \eqref{Delta} we have introduced the projection operators
\begin{equation}
\bra{p,\mathbf{k}}\mathcal{P}_n^\text{(pw/ew)}\ket{p',\mathbf{k}'}=k_z^n\bra{p,\mathbf{k}}\Pi^\text{(pw/ew)}\ket{p',\mathbf{k}'}
\end{equation}
where $\delta_{\phi\phi'}$ is the Kronecker delta and being $\Pi^\text{(pw)}$ [$\Pi^\text{(ew)}$] the projector on the propagative ($k<\omega/c$) [evanescent ($k>\omega/c$)] sector.

\section{FMM theory and grating scattering matrices}\label{SecFMM}

\begin{center}\begin{figure}[htb]
\includegraphics[width=8.5cm]{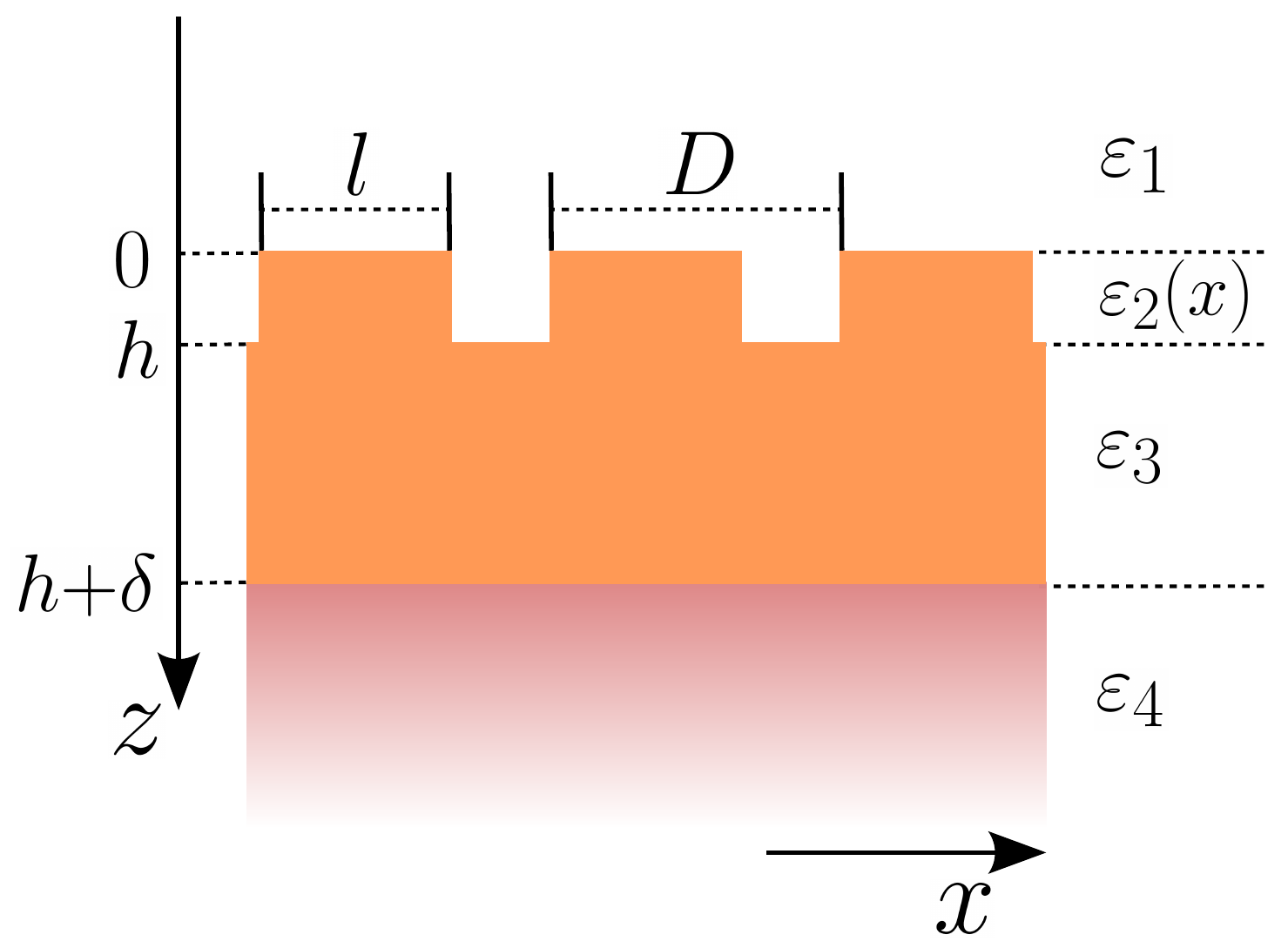}
\caption{(Color online) Geometry of the FMM calculation. We consider one grating with interface $z=0$, corrugation depth $h$, and underlying thickness $\delta$. This defines four zones (see text) with four (in general different) dielectric permittivities. The period is $D$ and the filling factor is defined as $f=l/D$.}\label{Fig2}\end{figure}\end{center}

In order to calculate the force, we now need to compute the reflection and transmission operators associated to a lamellar 1D grating. This will be achieved in the framework of the Fourier Modal Method (FMM) \cite{Kim12}. In the following, we implement this method for a grating of finite size along the $z$ axis (see Fig. \ref{Fig2}) in order to take into account finite-size effects on the Casimir-Lifshitz force. Moreover, we solve the scattering problem directly in TE and TM components, in order to be coherent with the formalism presented in Sec. \ref{SecIntro}.

Let us consider a system composed of a grating like the one in Fig. \ref{Fig2}. The space is divided in four zones: zone 1 ($z<0$), zone 2 ($0<z<h$), zone 3 ($h<z<h+\delta$) and zone 4 ($z>h+\delta$). While zones 1, 3 and 4 are homogeneous with dielectric permittivities $\varepsilon_i(\omega)$ ($i=1,3,4$), zone 2 represents the grating, with a dielectric function $\varepsilon_2(x,\omega)$, periodic in $x$ with period $D$. In each zone, every physical quantity is independent of $y$.

We first decompose the electric field in any zone with respect to frequency (only positive frequencies will be used):
\begin{equation}
\mathbf{E}^{(i)}(\mathbf{R},t)=2\Rea\Bigl[\int_0^{+\infty}\frac{d\omega}{2\pi}e^{-i\omega t}\mathbf{E}^{(i)}(\mathbf{R},\omega)\Bigr].
\end{equation}
In virtue of the translational invariance of our system along the $y$ axis and of the periodicity along the $x$ axis, we will employ a Fourier decomposition of any $x$-dependent quantity. As a consequence, the wavevector component $k_x$ will be replaced by a new mode variable
\begin{equation}k_{x,n}=k_x+\frac{2\pi}{D}n,\end{equation}
with $k_x$ taking values in the first Brillouin zone $[-\pi/D,\pi/D]$ and $n$ assuming all integer values.

\subsection{Homogeneous media}

In any homogeneous zone, we can use a standard Rayleigh expansion for the component of the field at frequency $\omega$
\begin{align}\label{EField}
\mathbf{E}^{(i)}(\mathbf{R},\omega)=\sum_{p,\phi}&\int_{-\frac{\pi}{D}}^{\frac{\pi}{D}}\frac{dk_x}{2\pi}\sum_{n\in\amsbb{Z}}\int_{-\infty}^{+\infty}\frac{dk_y}{2\pi}\nonumber\\
&e^{i\mathbf{K}^{(i)\phi}_n\cdot\mathbf{R}}\,\hat{\bbm[\epsilon]}^{(i)\phi}_p(\mathbf{k}_n,\omega)E^{(i)\phi}_p(\mathbf{k}_n,\omega),
\end{align}
where the wavevectors are defined as ($n\in\amsbb{Z}$)
\begin{equation}
\mathbf{K}^{(i)\phi}_n=(\mathbf{k}_n,\phi k^{(i)}_{z,n}),\quad\mathbf{k}_n=(k_{x,n},k_y),
\end{equation}
and $k^{(i)}_{z,n}$ is the $z$ component of the wavevector inside each medium
\begin{equation}
k^{(i)}_{z,n}=\sqrt{\varepsilon_i(\omega)\frac{\omega^2}{c^2}-\mathbf{k}_n^2}.
\end{equation}
The unit polarization vectors appearing in Eq. \eqref{EField} are defined as
\begin{align}
\hat{\bbm[\epsilon]}_\TE^{(i)\phi}(\mathbf{k}_n,\omega)&=\frac{1}{k_n}(-k_y\hat{\mathbf{x}}+k_{x,n}\hat{\mathbf{y}}),\\
\hat{\bbm[\epsilon]}_\TM^{(i)\phi}(\mathbf{k}_n,\omega)&=\frac{c}{\omega\sqrt{\epsilon_i(\omega)}}(-k_n\hat{\mathbf{z}}+\phi k^{(i)}_{z,n}\hat{\mathbf{k}_n}).
\end{align}
For convenience, we assign from now on the following labels to the field amplitudes in the three homogeneous zones (the dependence on $p$, $\mathbf{k}_n$ and $\omega$ is implicit):
\begin{align}&E^{(1)+}=I, &E^{(1)-}=R,\nonumber\\
&E^{(3)+}=C, &E^{(3)-}=C',\\
&E^{(4)+}=T, &E^{(4)-}=I',\nonumber
\end{align}
where $I$, $R$ and $T$ represent the incoming, reflected and transmitted field amplitudes respectively. The amplitude $I'$ is associated to a possible incoming field coming from the other side of the body. Its presence guarantees both the full symmetry of the calculation and the possibility to derive at the same time the reflection and transmission operators $\mathcal{R}^\pm$ and $\mathcal{T}^\pm$.

The magnetic field in any zone can be easily deduced from Maxwell's equations and it reads
\begin{align}
\label{BField}\mathbf{B}^{(i)}(\mathbf{R},\omega)=&\frac{\sqrt{\varepsilon_{i}(\omega)}}{c}\sum_{p,\phi}\int_{-\frac{\pi}{D}}^{\frac{\pi}{D}}\frac{dk_x}{2\pi}\sum_{n\in\amsbb{Z}}\int_{-\infty}^{+\infty}\frac{dk_y}{2\pi}\nonumber\\
&e^{i\mathbf{K}^{(i)\phi}_n\cdot\mathbf{R}}\,(-1)^p\hat{\bbm[\epsilon]}^{(i)\phi}_{S(p)}(\mathbf{k}_n,\omega)E^{(i)\phi}_p(\mathbf{k}_n,\omega),
\end{align}
where the function $S$ is defined as $S(1)=2$ and $S(2)=1$.

\subsection{Periodic region}

We now move to the periodic region (zone 2) where we write an arbitrary frequency component of the field as
\begin{equation}\label{E2Field}\mathbf{E}^{(2)}(\mathbf{R},\omega)=\int_{-\frac{\pi}{D}}^{\frac{\pi}{D}}\frac{dk_x}{2\pi}\sum_{n\in\amsbb{Z}}\int_{-\infty}^{+\infty}\frac{dk_y}{2\pi}e^{i\mathbf{k}_n\cdot\mathbf{r}}\mathbf{E}^{(2)}(z,\mathbf{k}_n,\omega),\end{equation}
where $\mathbf{R}=(\mathbf{r},z)$.

We are now ready to write Maxwell's equations (for our system we have $\partial_t=-i\omega$)
\begin{align} \label{eq:maxwell1}
&\begin{cases}
  \partial_y E_z - \partial _z E_y =i\omega \mu_0 H_x = ik_0\widetilde{H}_x \\
  \partial_z E_x - \partial _x E_z =i\omega \mu_0 H_y = ik_0\widetilde{H}_y \\
  \partial_x E_y - \partial _y E_x =i\omega \mu_0 H_z = ik_0\widetilde{H}_z
\end{cases}\nonumber\\
&\begin{cases}
  \partial_y H_z - \partial _z H_y =-i\omega \varepsilon \varepsilon_0 E_x  \\
  \partial_z H_x - \partial _x H_z =-i\omega \varepsilon \varepsilon_0 E_y  \\
  \partial_x H_y - \partial _y H_x =-i\omega \varepsilon \varepsilon_0 E_z
\end{cases}
\end{align}
where we used $\omega=ck_0$, $\omega\mu_0=k_0Z_0$, $\omega\varepsilon_0=k_0/Z_0$, $Z_0=\sqrt{\mu_0/\varepsilon_0}$ and defined $\widetilde{H}_i=Z_0H_i$. From \eqref{eq:maxwell1} we can easily obtain
\begin{align}\label{Maxwell1}
  &\partial_z \begin{pmatrix} E_x \\[5pt] E_y  \end{pmatrix}\\
  &=  \begin{pmatrix}
    -\dfrac{i}{k_0}\partial_x\dfrac{1}{\varepsilon(x)}\partial_y & ik_0+\dfrac{i}{k_0}\partial_x \dfrac{1}{\varepsilon(x)}\partial_x \\
    -ik_0-\dfrac{i}{k_0}\partial_y \dfrac{1}{\varepsilon(x)}\partial_y & \dfrac{i}{k_0}\partial_y\dfrac{1}{\varepsilon(x)}\partial_x
  \end{pmatrix} \begin{pmatrix} \widetilde{H}_x \\[5pt] \widetilde{H}_y \end{pmatrix},\nonumber
\end{align}

\begin{align}\label{Maxwell2}
  &\partial_z \begin{pmatrix} \widetilde{H}_x \\[5pt] \widetilde{H}_y \end{pmatrix}\\
  &=
  \begin{pmatrix}
    \dfrac{i}{k_0}\partial_x\partial_y & -ik_0\varepsilon(x)-\dfrac{i}{k_0}\partial_x \partial_x \\
    ik_0\varepsilon(x)+\dfrac{i}{k_0}\partial_y \partial_y & -\dfrac{i}{k_0}\partial_y\partial_x
  \end{pmatrix} \begin{pmatrix} E_x \\[5pt] E_y  \end{pmatrix}.\nonumber
\end{align}
We now employ a Fourier factorization for the fields $E$ and $\widetilde{H}$. Correspondingly, the operator $\partial_y$ is replaced by $i\beta$, $\beta$ being a scalar, whereas the operator $\partial_x$ is replaced by $i\alpha$, where $\alpha=\diag(k_{x,n})_n$. These replacements allow us to rewrite Maxwell's equations of our system in a more compact form:
\begin{align}\label{Maxwell3}
  \partial_z \bm{\mathcal{E}} &=
  \begin{pmatrix}
    \dfrac{i\beta}{k_0}\alpha {\llbracket \varepsilon\rrbracket}^{-1} & ik_0\mathbb{1}-\dfrac{i\alpha}{k_0} {\llbracket\varepsilon\rrbracket}^{-1}\alpha \\
    -ik_0\mathbb{1}+\dfrac{i\beta^2}{k_0}{\llbracket\varepsilon\rrbracket}^{-1} & -\dfrac{i\beta}{k_0}{\llbracket\varepsilon\rrbracket}^{-1}\alpha
  \end{pmatrix} \widetilde{\bm{\mathcal{H}}}\nonumber\\
  &=\amsbb{F}\widetilde{\bm{\mathcal{H}}},
\end{align}
\begin{align}\label{Maxwell4}
  \partial_z \widetilde{\bm{\mathcal{H}}} &=
  \begin{pmatrix}
    -\dfrac{i\beta}{k_0}\alpha  & -ik_0 {\llbracket\varepsilon\rrbracket}+\dfrac{i\alpha^2}{k_0} \\[8pt]
    ik_0{\bigg\llbracket\dfrac{1}{\varepsilon}\bigg\rrbracket}^{\!-1}\!\!\!-\dfrac{i\beta^2}{k_0} & \dfrac{i\beta}{k_0}\alpha
  \end{pmatrix} \bm{\mathcal{E}}\nonumber\\
  &=\amsbb{G}\bm{\mathcal{E}},
\end{align}
where for an arbitrary field $\mathbf{U}$ we have introduced the decomposition
\begin{equation}\label{DefUbm}\bm{\mathcal{U}}=\bigl({\{U_x(z,\mathbf{k}_n,\omega)\}}_n,{\{U_y(z,\mathbf{k}_n,\omega)\}}_n\bigr)^T,\end{equation}
gathering $x$ and $y$ components and denoting with ${\{\dots\}}_n$ a set of scattering orders. We have also introduced the Toeplitz matrix $\llbracket a\rrbracket$, defined by the relation $\llbracket a\rrbracket_{ij}=a_{i-j}$, $a_n$ being the $n$-th Fourier component of $a$. We remark that going from Eqs. \eqref{Maxwell1}-\eqref{Maxwell2} to Eqs. \eqref{Maxwell3}-\eqref{Maxwell4} we have used the modified factorization rule introduced in \cite{GranetJOptSocAmA96}.

Of course, in order to exploit numerically the FMM, a truncation has to be made, limiting the number of diffraction orders taken into account. For a given truncation $M$, this corresponds to keeping $2M+1$ scattering orders
\begin{equation}{\{A_n\}}_n=\bigl(A_{-M},\dots,A_{M}\bigr),\end{equation}
and the size of the corresponding column vector $\bm{\mathcal{U}}$ is thus $2(2M+1)$. Based on this truncation, we obtain
\begin{align}\label{MaxComp}
  \partial^2_z\bm{\mathcal{E}}=\amsbb{F}\amsbb{G}\bm{\mathcal{E}}=\amsbb{PD}^2\amsbb{P}^{-1}\bm{\mathcal{E}},
\end{align}
where $\amsbb{P}$ and $\amsbb{D}^2$ are respectively the eigenvectors and eigenvalues $2(2M+1)\times2(2M+1)$ matrices of the matrix $\amsbb{F}\amsbb{G}$
\begin{align}
  \amsbb{P}=\begin{pmatrix} \amsbb{P}^{(11)} & \amsbb{P}^{(12)} \\
  \amsbb{P}^{(21)} & \amsbb{P}^{(22)} \end{pmatrix}, \qquad   \amsbb{D}=\begin{pmatrix} \amsbb{D}^{(11)} & \mathbb{0} \\
  \mathbb{0} & \amsbb{D}^{(22)} \end{pmatrix}.
\end{align}
Then, from Eqs. \eqref{Maxwell3} and \eqref{MaxComp}, we obtain that fields are
\begin{align}\label{eq:Fields_compact}
  \begin{cases}
    \bm{\mathcal{E}}(z)=\amsbb{P}\Bigl(e^{\amsbb{D}z}\bm{\mathcal{A}}+e^{\amsbb{-D}z}\bm{\mathcal{B}}\Bigr)\\
    \widetilde{\bm{\mathcal{H}}}(z)=\amsbb{P}'\Bigl(e^{\amsbb{D}z}\bm{\mathcal{A}}-e^{\amsbb{-D}z}\bm{\mathcal{B}}\Bigr)
  \end{cases}
\end{align}
$\bm{\mathcal{A}}$ and $\bm{\mathcal{B}}$ being arbitrary constant vectors, and where $\amsbb{P}'=\amsbb{F}^{-1}\amsbb{P}\amsbb{D}$.

\subsection{Boundary conditions}

Based on the knowledge of the electric and magnetic fields in the four regions, we can now impose the continuity of the $x$ and $y$ components of both fields at the three interfaces $z=0$, $z=h$ and $z=h+\delta$. In the following boundary conditions the values of $k_x$, $k_y$ and $\omega$ are given. Exploiting this fact we use the generic simplified expression $A_{p,n}$ to refer to the amplitude $A_p(\mathbf{k}_n,\omega)$. Before proceeding in the calculation, we introduce an additional phase factor in the expression of the fields in zones 3 and 4. In particular, in zone 3 we replace $\exp[ik_z^{(i)\phi}z]$ with $\exp[ik_z^{(i)\phi}(z-h)]$, while in zone 4 we replace $\exp[ik_z^{(i)\phi}z]$ with $\exp[ik_z^{(i)\phi}(z-h-\delta)]$. These factors make the calculation easier and can be simply recovered at the end. At the first interface $z=0$ we have for the $x$ and $y$ components of the electric field (repeated indices are implicitly summed over)
\begin{widetext}
\begin{equation}\label{SistemaIn}
\begin{pmatrix}-\frac{k_y}{k_n}\big(I_{1,n}+R_{1,n} \big)+ \frac{c}{\sqrt{\varepsilon_1}\omega}k_{z,n}^{(1)}\frac{k_{x,n}}{k_n}(I_{2,n}-R_{2,n})\\[5pt]
\frac{k_{x,n}}{k_n}(I_{1,n}+R_{1,n})+\frac{c}{\sqrt{\varepsilon_1}\omega}k_{z,n}^{(1)}\frac{k_y}{k_n}(I_{2,n}-R_{2,n})\end{pmatrix}=
\begin{pmatrix}\amsbb{P}^{(11)}_{nm}(A_{x,m}+B_{x,m})+\amsbb{P}^{(12)}_{nm}(A_{y,m}+B_{y,m})\\[5pt]
\amsbb{P}^{(21)}_{nm}(A_{x,m}+B_{x,m})+\amsbb{P}^{(22)}_{nm}(A_{y,m}+B_{y,m}) \end{pmatrix},
\end{equation}
while for the magnetic field we get
\begin{equation}
\begin{pmatrix} -\frac{c}{\omega} k_{z,n}^{(1)} \frac{k_{x,n}}{k_n} \big(I_{1,n} - R_{1,n} \big) -\sqrt{\varepsilon_1}\frac{k_y}{k_n} \big(I_{2,n}+R_{2,n} \big) \\[5pt]
-\frac{c}{\omega} k_{z,n}^{(1)} \frac{k_y}{k_n} \big( I_{1,n} -R_{1,n} \big) + \sqrt{\varepsilon_1}\frac{k_{x,n}}{k_n} \big(I_{2,n}+R_{2,n} \big)
\end{pmatrix}=\begin{pmatrix}{\amsbb{P}'}^{(11)}_{nm}(A_{x,m}-B_{x,m})+{\amsbb{P}'}^{(12)}_{nm}(A_{y,m}-B_{y,m})\\[5pt]
{\amsbb{P}'}^{(21)}_{nm}(A_{x,m}-B_{x,m})+{\amsbb{P}'}^{(22)}_{nm}(A_{y,m}-B_{y,m})\end{pmatrix}.
\end{equation}
The boundary conditions at $z=h$ give us the following equations for the electric field
\begin{align}
&\begin{pmatrix}-\frac{k_y}{k_n}(C_{1,n}+C'_{1,n})+\frac{c}{\sqrt{\varepsilon_3}\omega}k_{z,n}^{(3)}\frac{k_{x,n}}{k_n}({C}_{2,n}-C'_{2,n})\\[5pt]
\frac{k_{x,n}}{k_n}({C}_{1,n}+C'_{1,n})+\frac{c}{\sqrt{\varepsilon_3}\omega}k_{z,n}^{(3)}\frac{k_y}{k_n}(C_{2,n}-C'_{2,n})
\end{pmatrix}\nonumber\\
&=\begin{pmatrix}\amsbb{P}^{(11)}_{nm}\Big(e^{\amsbb{D}^{(11)}_{mm}h}A_{x,m}+e^{-\amsbb{D}^{(11)}_{mm}h}B_{x,m}\Big)+\amsbb{P}^{(12)}_{nm}\Big(e^{\amsbb{D}^{(22)}_{mm}h}\,A_{y,m}+e^{-\amsbb{D}^{(22)}_{m}h}B_{y,m}\Big)\\[5pt]
\amsbb{P}^{(21)}_{nm}\Big(e^{\amsbb{D}^{(11)}_{mm}h}\,A_{x,m}+e^{-\amsbb{D}^{(11)}_{mm}h}B_{x,m}\Big)+\amsbb{P}^{(22)}_{nm}\Big(e^{\amsbb{D}^{(22)}_{mm}h}A_{y,m}+e^{-\amsbb{D}^{(22)}_{mm}h}B_{y,m}\Big)\end{pmatrix},
\end{align}
and the following ones for the magnetic field
\begin{align}
& \begin{pmatrix}-\frac{c}{\omega}k_{z,n}^{(3)}\frac{k_{x,n}}{k_n}(C_{1,n}-C'_{1,n})-\sqrt{\varepsilon_3}\frac{k_y}{k_n}(C_{2,n}+C'_{2,n})\\[5pt]
-\frac{c}{\omega} k_{z,n}^{(3)}\frac{k_y}{k_n}(C_{1,n}-C'_{1,n})+\sqrt{\varepsilon_3}\frac{k_{x,n}}{k_n}(C_{2,n}+C'_{2,n})
\end{pmatrix}\nonumber\\
&=\begin{pmatrix} {\amsbb{P}'}^{(11)}_{nm}\Big(e^{\amsbb{D}^{(11)}_{mm}h}\,A_{x,m}-e^{-\amsbb{D}^{(11)}_{mm}h}\,B_{x,m}\Big)+{\amsbb{P}'}^{(12)}_{nm}\Big(e^{\amsbb{D}^{(22)}_{mm}h}\,A_{y,m}-e^{-\amsbb{D}^{(22)}_{mm}h}\,B_{y,m}\Big)\\[5pt]
{\amsbb{P}'}^{(21)}_{nm}\Big(e^{\amsbb{D}^{(11)}_{mm}h}\,A_{x,m}-e^{-\amsbb{D}^{(11)}_{mm}h}\,B_{x,m}\Big)+{\amsbb{P}'}^{(22)}_{nm}\Big(e^{\amsbb{D}^{(22)}_{mm}h}\,A_{y,m}-e^{-\amsbb{D}^{(22)}_{mm}h}\,B_{y,m}\Big)\end{pmatrix}.\end{align}
Finally, the boundary conditions at $z=h+\delta$ read
\begin{align}
&\begin{pmatrix}-\frac{k_y}{k_n}\big(e^{i k_{z,n}^{(3)}\delta}\,C_{1,n}+e^{-i k_{z,n}^{(3)}\delta}\,C'_{1,n}\big)+\frac{c}{\sqrt{\varepsilon_3}\omega}k_{z,n}^{(3)}\frac{k_{x,n}}{k_n}\big(e^{i k_{z,n}^{(3)}\delta}\,{C}_{2,n}-e^{-i k_{z,n}^{(3)}\delta}\,C'_{2,n}\big)\\[5pt]
\frac{k_{x,n}}{k_n}\big(e^{i k_{z,n}^{(3)}\delta}\,{C}_{1,n}+e^{-i k_{z,n}^{(3)}\delta}\,C'_{1,n}\big)+\frac{c}{\sqrt{\varepsilon_3}\omega}k_{z,n}^{(3)}\frac{k_y}{k_n}\big(e^{i k_{z,n}^{(3)}\delta}\,C_{2,n}-e^{-i k_{z,n}^{(3)}\delta}\,C'_{2,n}\big)
\end{pmatrix}\nonumber\\
&=\begin{pmatrix}-\frac{k_y}{k_n}(T_{1,n}+I'_{1,n})+\frac{c}{\sqrt{\varepsilon_4}\omega}k_{z,n}^{(4)}\frac{k_{x,n}}{k_n}(T_{2,n}-I'_{2,n})\\[5pt]
\frac{k_{x,n}}{k_n}(T_{1,n}+I'_{1,n})+\frac{c}{\sqrt{\varepsilon_4}\omega}k_{z,n}^{(4)}\frac{k_y}{k_n}\big(T_{2,n}-I'_{2,n})
\end{pmatrix},
\end{align}
and the ones for the magnetic field are given by
\begin{align}\label{SistemaFin}
&\begin{pmatrix}-\frac{c}{\omega}k_{z,n}^{(3)}\frac{k_{x,n}}{k_n}\big(e^{i k_{z,n}^{(3)}\delta}\,C_{1,n}-e^{-i k_{z,n}^{(3)}\delta}\,C'_{1,n}\big)-\sqrt{\varepsilon_3}\frac{k_y}{k_n}\big(e^{i k_{z,n}^{(3)}\delta}\, C_{2,n}+e^{-i k_{z,n}^{(3)}\delta}\,C'_{2,n}\big)\\[5pt]
-\frac{c}{\omega} k_{z,n}^{(3)}\frac{k_y}{k_n}\big(e^{i k_{z,n}^{(3)}\delta}\,C_{1,n}-e^{-i k_{z,n}^{(3)}\delta}\,C'_{1,n}\big)+\sqrt{\varepsilon_3}\frac{k_{x,n}}{k_n}\big(e^{i k_{z,n}^{(3)}\delta}\,C_{2,n}+e^{-i k_{z,n}^{(3)}\delta}\,C'_{2,n}\big)
\end{pmatrix}\nonumber\\
&=\begin{pmatrix}-\frac{c}{\omega}k_{z,n}^{(4)}\frac{k_{x,n}}{k_n}(T_{1,n}-I'_{1,n})-\sqrt{\varepsilon_4}\frac{k_y}{k_n}(T_{2,n}+I'_{2,n})\\[5pt]
-\frac{c}{\omega} k_{z,n}^{(4)}\frac{k_y}{k_n}(T_{1,n}-I'_{1,n})+\sqrt{\varepsilon_4}\frac{k_{x,n}}{k_n}(T_{2,n}+I'_{2,n})\end{pmatrix}.
\end{align}
\end{widetext}

\subsection{Scattering matrices}

In the following, we are going to cast Eqs. \eqref{SistemaIn}-\eqref{SistemaFin} under the form
\begin{align}\label{SistemaComp}
\begin{pmatrix}\bm{\mathcal{R}}\\\bm{\mathcal{A}}\end{pmatrix}=\amsbb{S}_1\begin{pmatrix}\bm{\mathcal{I}}\\\bm{\mathcal{B}}\end{pmatrix},\hspace{.2cm}\begin{pmatrix}\bm{\mathcal{B}}\\\bm{\mathcal{C}}\end{pmatrix}=\amsbb{S}_2\begin{pmatrix}\bm{\mathcal{A}}\\\bm{\mathcal{C}}'\end{pmatrix},\hspace{.2cm}\begin{pmatrix}\bm{\mathcal{C}}'\\\bm{\mathcal{T}}\end{pmatrix}=\amsbb{S}_3\begin{pmatrix}\bm{\mathcal{C}}\\\bm{\mathcal{I}}'\end{pmatrix}.
\end{align}
The column vectors $\bm{\mathcal{A}}$ and $\bm{\mathcal{B}}$ appearing in this equation gather two vectors defined as in Eq. \eqref{DefUbm}. On the contrary, all the six other column vectors gather the two polarizations of the field under the form
\begin{equation}\bm{\mathcal{V}}=\bigl({\{V_1(z,\mathbf{k}_n,\omega)\}}_n,{\{V_2(z,\mathbf{k}_n,\omega)\}}_n\bigr)^T.\end{equation}
The system of equations \eqref{SistemaComp} has to be solved for the unknowns $\bm{\mathcal{R}}$, $\bm{\mathcal{T}}$, $\bm{\mathcal{A}}$, $\bm{\mathcal{B}}$, $\bm{\mathcal{C}}$, and $\bm{\mathcal{C}}'$. The expression of $\bm{\mathcal{R}}$ and $\bm{\mathcal{T}}$ as a function of $\bm{\mathcal{I}}$ and $\bm{\mathcal{I}}'$ will provide us the desired reflection and transmission operators. The fact that for $\bm{\mathcal{A}}$ and $\bm{\mathcal{B}}$ we solve in cartesian components and not in polarization is not an issue since these appear as mute variables not participating to the scattering operators.

The explicit expression of the $\amsbb{S}$ matrices appearing in \eqref{SistemaComp} can be obtained by means of algebraic manipulation of Eqs. \eqref{SistemaIn}-\eqref{SistemaFin}. The final result is
\begin{align}
\amsbb{S}_1=
\begin{pmatrix}
\amsbb{K}'_1 & -\amsbb{P} \\
\amsbb{L}'_1 & -\amsbb{P}'
\end{pmatrix}^{-1}
\begin{pmatrix}
\amsbb{K}_1 & \amsbb{P} \\
\amsbb{L}_1 & -\amsbb{P}'
\end{pmatrix},
\end{align}
\begin{align}
\amsbb{S}_2=
\begin{pmatrix}
\bbsigma^{(2)}_h & \mathbb{0} \\
\mathbb{0} & \mathbb{1}
\end{pmatrix}
\begin{pmatrix}
-\amsbb{P} & -\amsbb{K}_3\\
\amsbb{P}' & -\amsbb{L}_3
\end{pmatrix}^{-1}
\begin{pmatrix}
\amsbb{P} & -\amsbb{K}'_3\\
\amsbb{P}' & -\amsbb{L}'_3
\end{pmatrix}
\begin{pmatrix}
\bbsigma^{(2)}_h & \mathbb{0} \\
\mathbb{0} & \mathbb{1}
\end{pmatrix},
\end{align}
\begin{align}
\amsbb{S}_3=
\begin{pmatrix}
\bbsigma^{(3)}_\delta & \mathbb{0} \\
\mathbb{0} & \mathbb{1}
\end{pmatrix}
\begin{pmatrix}
\amsbb{K}'_3 & \amsbb{K}_4\\
\amsbb{L}'_3 & \amsbb{L}_4
\end{pmatrix}^{-1}
\begin{pmatrix}
\amsbb{K}_3 & \amsbb{K}'_4\\
\amsbb{L}_3 & \amsbb{L}'_4
\end{pmatrix}
\begin{pmatrix}
\bbsigma^{(3)}_\delta & \mathbb{0} \\
\mathbb{0} & \mathbb{1}
\end{pmatrix}.
\end{align}
In these expressions we have defined
\begin{align}
&\amsbb{K}'_i=
\begin{pmatrix}
-\amsbb{A}_y & -\amsbb{B}_{x,i}\\
\amsbb{A}_x & -\amsbb{B}_{y,i}
\end{pmatrix},\,
&\amsbb{L}'_i=
\sqrt{\varepsilon_i}\begin{pmatrix}
\amsbb{B}_{x,i} & -\amsbb{A}_y\\
\amsbb{B}_{y,i} & \amsbb{A}_x
\end{pmatrix},\nonumber\\
&\amsbb{K}_i=
\begin{pmatrix}
\amsbb{A}_y & -\amsbb{B}_{x,i}\\
-\amsbb{A}_x & -\amsbb{B}_{y,i}
\end{pmatrix},\,
&\amsbb{L}_i=
\sqrt{\varepsilon_i}\begin{pmatrix}
\amsbb{B}_{x,i} & \amsbb{A}_y\\
\amsbb{B}_{y,i} & -\amsbb{A}_x
\end{pmatrix},
\end{align}
where
\begin{align}
\amsbb{A}_x&=\diag\Big(\frac{k_{x,n}}{k_n}\Big)_n,\quad\amsbb{A}_y=\diag\Big(\frac{k_y}{k_n}\Big)_n,\nonumber\\
\amsbb{B}_{x,i}&=\frac{c}{\sqrt{\varepsilon_i}\omega}\diag\Big(\frac{k_{x,n}}{k_n}k_{z,n}^{(i)}\Big)_n,\\
\amsbb{B}_{y,i}&=\frac{c}{\sqrt{\varepsilon_i}\omega}\diag\Big(\frac{k_y}{k_n}k_{z,n}^{(i)}\Big)_n.\nonumber
\end{align}
The symbol $\diag(a_n)_n$ denotes a $(2M+1)\times(2M+1)$ diagonal matrix having diagonal elements $a_{-M}$, $a_{-M+1}$, \dots, $a_M$. We have also defined the square matrices of dimension $2(2M+1)$
\begin{align}
&\bbsigma^{(2)}_h\equiv e^{\amsbb{D}h}=\begin{pmatrix} e^{\amsbb{D}^{(11)}h} & \mathbb{0} \\
\mathbb{0} & e^{\amsbb{D}^{(22)}h} \end{pmatrix},\\
&\bbsigma^{(3)}_\delta\equiv\begin{pmatrix}\diag(e^{ik_{z,n}^{(3)}\delta})_n & \mathbb{0} \\
\mathbb{0} & \diag(e^{ik_{z,n}^{(3)}\delta})_n\end{pmatrix}.
\end{align}

Using \eqref{SistemaComp} we obtain the final result
\begin{align}\label{RisS}
\begin{pmatrix}\bm{\mathcal{R}}\\\bm{\mathcal{T}}\end{pmatrix}=\amsbb{S}\begin{pmatrix}\bm{\mathcal{I}}\\\bm{\mathcal{I}}'\end{pmatrix},
\end{align}
where
\begin{align}
\amsbb{S}=\amsbb{S}_1\circledast\amsbb{S}_2\circledast\amsbb{S}_3,
\end{align}
having introduced the associative operation $\amsbb{A}=\amsbb{B}\circledast\amsbb{C}$, which for three square matrices $\amsbb{A}$, $\amsbb{B}$ and $\amsbb{C}$ of dimension $4(2M+1)$ is defined as
\begin{align}
\amsbb{A}_{11}&=\amsbb{B}_{11}+\amsbb{B}_{12}(\mathbb{1}-\amsbb{C}_{11}\amsbb{B}_{22})^{-1}\amsbb{C}_{11}\amsbb{B}_{21},\\
\amsbb{A}_{12}&=\amsbb{B}_{12}(\mathbb{1}-\amsbb{C}_{11}\amsbb{B}_{22})^{-1}\amsbb{C}_{12},\\
\amsbb{A}_{21}&=\amsbb{C}_{21}(\mathbb{1}-\amsbb{B}_{22}\amsbb{C}_{11})^{-1}\amsbb{B}_{21},\\
\amsbb{A}_{22}&=\amsbb{C}_{22}+\amsbb{C}_{21}(\mathbb{1}-\amsbb{B}_{22}\amsbb{C}_{11})^{-1}\amsbb{B}_{22}\amsbb{C}_{12},
\end{align}
where each matrix have been decomposed in four square blocks of dimension $2(2M+1)$.

Equation \eqref{RisS} allows to identify the four blocks of $\amsbb{S}$ as the reflection and transmission operators associated to the two sides of the grating. For example, the block $\amsbb{S}_{11}$ is the coefficient linking the reflected amplitudes $\bm{\mathcal{R}}$ to the incident ones $\bm{\mathcal{I}}$: it then coincides with the reflection operator $\mathcal{R}^-$ for a wave impinging on the grating of Fig. \eqref{Fig2} from $z<0$. By analog reasoning, we write the full $\amsbb{S}$ matrix as
\begin{align}
&\amsbb{S}=
\begin{pmatrix}
\mathcal{R}^- & \mathcal{T}^-\\
\mathcal{T}^+ & \mathcal{R}^+
\end{pmatrix}.
\end{align}

\subsection{Two lamellar gratings}

We now need to calculate the reflection and transmission operators associated to the two gratings represented in Fig. \ref{Fig1}. As far as grating 1 is concerned, the problem we need to solve is exactly the one presented in this Section, with the appropriate values of the geometrical parameters. Concerning grating 2, we need to take into account the fact that its interface is the plane $z=d$ and not $z=0$. The modification of the scattering operators with respect to translations has been discussed in \cite{MessinaPRA11}. Based on these results, and using the mode expansion used in this work, the $\mathcal{R}_2^-$ operator of grating 2 can be expressed as a function of the $\widetilde{\mathcal{R}}_2^-$ derived from FMM as
\begin{align}
&\bra{p,\mathbf{k},n,\omega}\mathcal{R}_2^-\ket{p',\mathbf{k}',n',\omega'}\\
&=\exp[i(k_{z,n}+k'_{z,n'})d]\bra{p,\mathbf{k},n,\omega}\widetilde{\mathcal{R}}_2^-\ket{p',\mathbf{k}',n',\omega'}.\nonumber
\end{align}
As we will show in the next Section, this operator is the only one associated to grating 2 appearing in the expression of the force for our configuration.

\section{Numerical results}\label{SecNum}

In this Section we will present a numerical application concerning the force between two different gratings. Being both gratings infinite in the $xy$ plane, we actually calculate the pressure acting on any of them, as discussed in the case of two slabs in \cite{MessinaPRA11}. In the first configuration we have chosen both gratings to have period $D=1\,\mu$m, corrugation depth $h=1\,\mu$m and filling factor $f=0.5$. As shown in Fig. \ref{Fig1}, the transition points of the two gratings are aligned, i.e. there is no shift along the $x$ axis. Grating 1 is made of Fused Silica (SiO$_2$) and has thickness $\delta_1=10\,\mu$m, while grating 2 is made of Silicon and has infinite thickness. In order to take into account this point we have imposed $\varepsilon_3=\varepsilon_4$ in the FMM relative to grating 2 (see Sec. \ref{SecFMM}) and removed in Eq. \eqref{Delta} all the terms proportional to the transmission operators of body 2. Physically, this can be explained by observing that because of the infinite thickness all the radiation coming from the upper side of body 2 is absorbed and does not reach the cavity between the gratings. Both Silicon and Fused Silica have been described by means of optical data taken from \cite{Palik98}.

\subsection{The issue of convergence}

As anticipated in Sec. \ref{SecFMM}, the numerical use of FMM demands to choice of a truncation order, problem that will be addressed in this Section. We noted before that by choosing a truncation order $M$ in the FMM we obtain as a result reflection operators which are square matrices of dimension $2(2M+1)$, that is two polarizations times $2M+1$ diffraction orders. Their typical structure is thus
\begingroup
\renewcommand*{\arraystretch}{1.5}
\vspace{-0.4cm}
\begin{equation}
\begin{array}{lc}
 & \begin{array}{cc}\text{TE}\quad\quad&\quad\quad\text{TM}\end{array}\\
\begin{array}{c}\text{TE}\\\text{TM}\end{array}  & \left(\begin{array}{c|c}A_{1,1}[n,n'] & A_{1,2}[n,n']\\\hline A_{2,1}[n,n'] & A_{2,2}[n,n']\end{array}\right)\end{array},\end{equation}
\endgroup
where each block $A_{i,j}[n,n']$ is a $(2M+1)\times(2M+1)$ matrix, the indices $n$ and $n'$ running from $-M$ to $M$.

It is worth stressing that, for a given $M$, only the elements closer to the center of each block of the matrix (i.e. close to $n=0$ for each couple of polarizations) are at convergence. Thus, for a given $m$, we can increase the value of $M$ starting from $M=m$ in order to extract a $2(2m+1)\times2(2m+1)$ ($m<M$) scattering operator whose elements are at convergence with a given accuracy (in our case of the order of one percent). The operators obtained following this procedure can be used to compute the force using Eqs. \eqref{Feq} and \eqref{Delta}. Since these equations imply a trace containing also a sum over the diffraction orders $n$, the series has to be replaced with a finite sum from $-\bar{m}$ to $\bar{m}$. The value of $\bar{m}$ has to be found by imposing the convergence of the series at a chosen accuracy. Also in this case, we required an accuracy smaller than one percent.

The calculation of the pressure at a given distance requires the evaluation of the traces \eqref{Feq} and \eqref{Delta} at several different values of the wavevector $\mathbf{k}$ and the frequency $\omega$, in order to reach the convergence on the integral on the three variables. We have observed that a single calculation of the trace requires values of $\bar{m}$ of the order of 2 (with peaks going up to 7) and corresponding values of $M$ of the order of 5 (with peaks around 20). A single value of the pressure required a computation time of the order of 16 hours on three 3\,GHz CPUs.

\subsection{Casimir-Lifshitz force OTE between\\two different gratings}

\begin{center}\begin{figure}[htb]
\includegraphics[width=8cm]{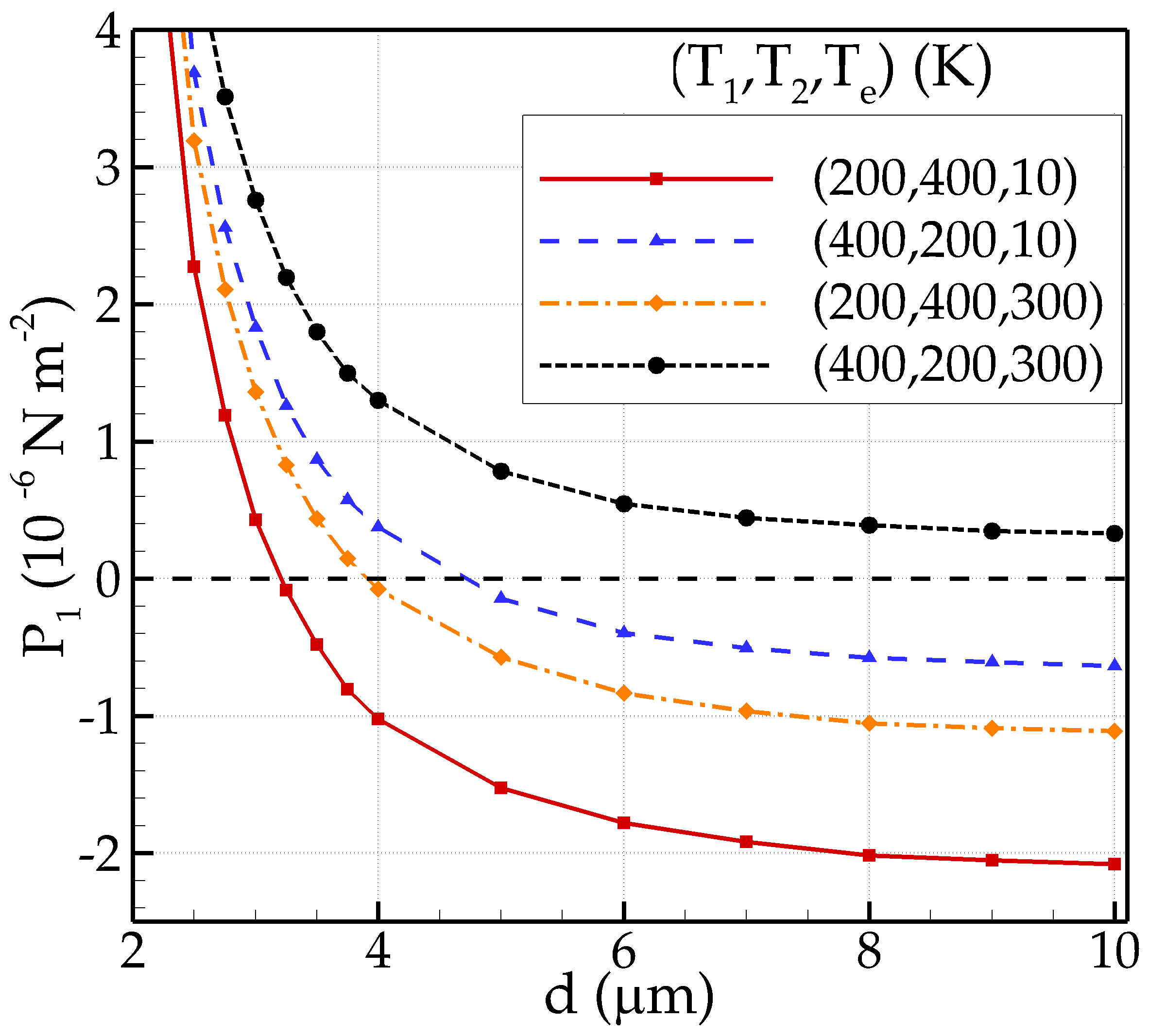}
\caption{(Color online) Pressure acting on grating 1 (made of Fused Silica, having $h_1=1\,\mu$m, $\delta_1=10\,\mu$m, $D=1\,\mu$m and $f_1=0.5$) in front of grating 2 (made of Silicon, having $h_2=1\,\mu$m, infinite thickness, $D=1\,\mu$m and $f_2=0.5$) as a function of distance $d$. The four curves correspond to different choices of the three temperatures $(T_1,T_2,T_\text{e})$ (see legend).}\label{Fig3}\end{figure}\end{center}

In the configuration described above, we have calculated the pressure acting on grating 1. To point out the features of our OTE configuration we present in Fig. \ref{Fig3} the pressure as a function of distance for different sets of the temperatures $(T_1,T_2,T_\text{e})$.

We clearly see that the modification of the three temperatures strongly affects the value of the force. In particular, three of the four curves show a transition from an attractive to a repulsive behavior, not realizable at thermal equilibrium for this configuration. This qualitative difference is a well-known consequence of the absence of thermal equilibrium and it has already been predicted in the case of two parallel slabs \cite{AntezzaPRA08,MessinaPRA11}. We stress that the transition point between attraction and repulsion is a function of the temperatures. For the values chosen, it roughly varies from 3 to 5\,$\mu$m.

\begin{center}\begin{figure}[htb]
\includegraphics[width=8cm]{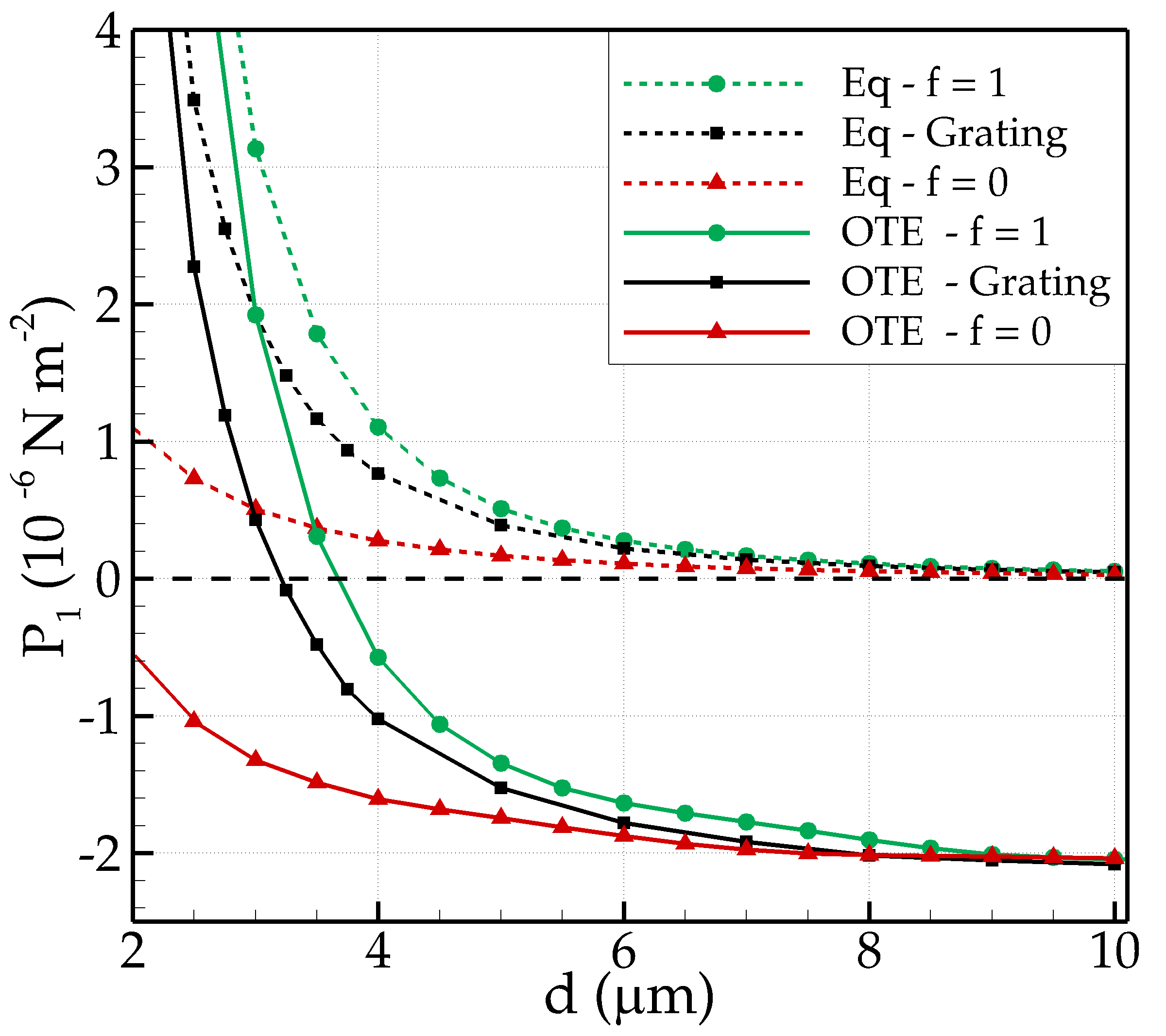}
\caption{(Color online) Non-equilibrium (OTE) pressure [$(T_1,T_2,T_\text{e})=(200,400,10)\,$K, solid lines] compared to equilibrium pressure ($T=200\,$K, dashed lines) for two gratings (black squares), and two slab-slab configurations corresponding to filled gratings ($f=1$, green circles) and an empty ones ($f=0$, red triangles).}\label{Fig4}\end{figure}\end{center}

To underline even more the richness of our OTE configuration, we focus on the temperatures $(T_1,T_2,T_\text{e})=(200,400,10)\,$K and compare the pressure to its equivalent at thermal equilibrium at the temperature of body 1, i.e. $T_1=200\,$K. This comparison is presented in Fig. \ref{Fig4}. In the same figure we also plot the pressure, both at and out of thermal equilibrium, for filling factors $f_1=f_2=1$ (corresponding to \emph{filled} gratings, that is a 11\,$\mu$m-thick SiO$_2$ slab at distance $d$ from an infinite Si slab) and for $f_1=f_2=0$ (corresponding to \emph{empty} gratings, that is a 10\,$\mu$m-thick SiO$_2$ slab at distance $d+2\,\mu$m from an infinite Si slab).

Apart from the transition to a repulsive behavior, this figure shows that the pressure in presence of a grating always lies between the two results corresponding to filled and empty ones. Finally, a comparison between Figs. \ref{Fig3} and \ref{Fig4} shows that the asymptotic value of the pressure can be tuned by varying the temperatures to values comparable (apart from their sign) to the pressure at thermal equilibrium at much smaller distances, of the order of 3\,$\mu$m.

To conclude this Section, we compare the grating-grating pressure obtained using FMM to the result coming from the PFA (Proximity Force Approximation), typically used to deal with complex geometries such as sphere-plane and nanostructured surfaces. In the case of two aligned gratings with equal filling factors $f_1=f_2=f$ it reduces to the following weighted sum of the pressures of simple slab-slab configurations \cite{BaoPRL10,LussangePRB12}:
\begin{align}\label{PPFA}P_{1,\text{PFA}}(d)&=fP_1^{\text{(ss)}}(\delta_1,\delta_2,d)\\
&\,+(1-f)P_1^{\text{(ss)}}(\delta_1-h_1,\delta_2-h_2,d+h_1+h_2),\nonumber\end{align}
where $P_1^{\text{(ss)}}(\delta_1,\delta_2,d)$ is the pressure acting on a $\delta_1$-thick slab at a distance $d$ from a $\delta_2$-thick slab.

\begin{center}\begin{figure}[htb]
\includegraphics[width=8cm]{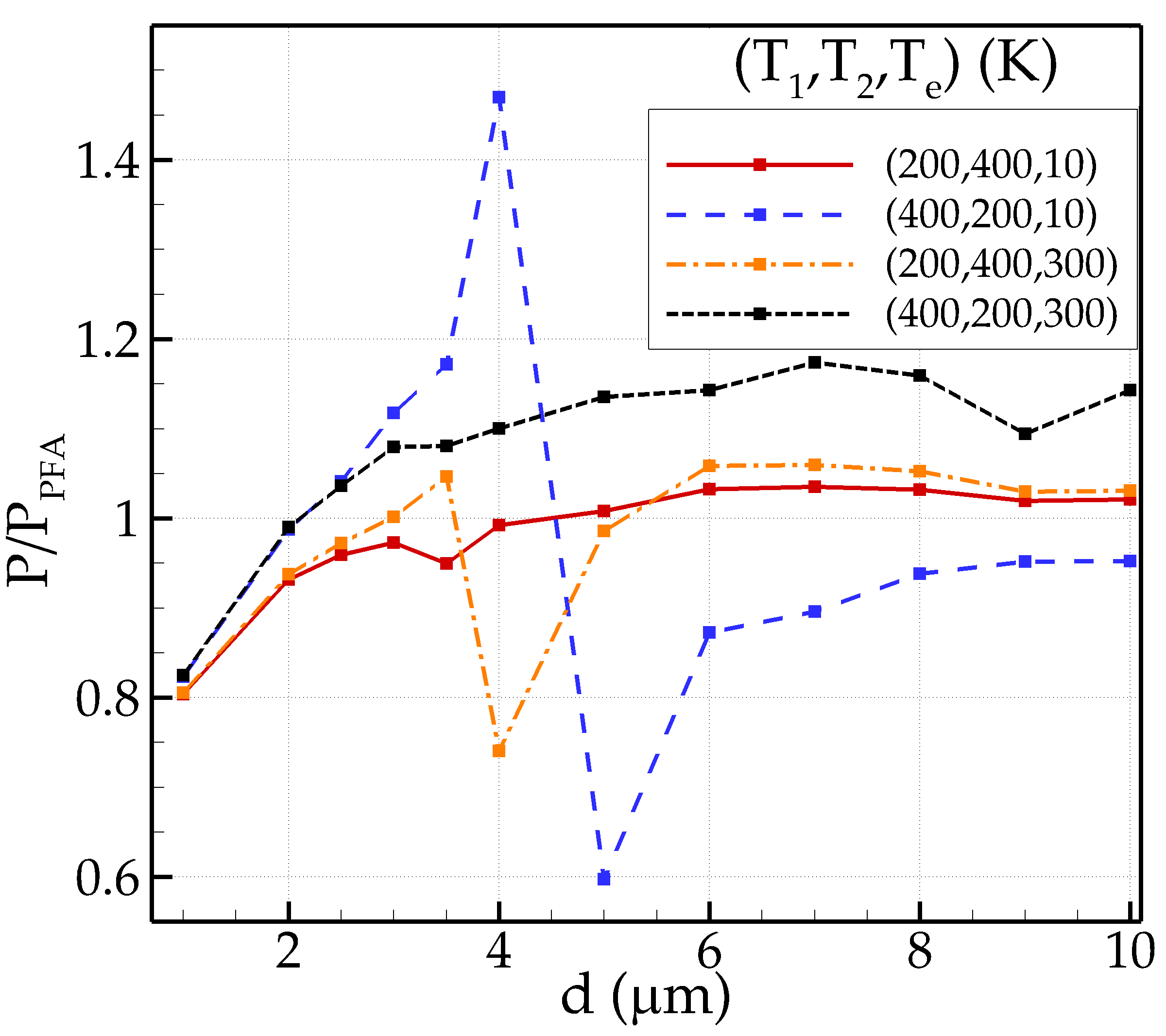}
\caption{(Color online) Ratio between the exact pressure and the PFA counterpart (see Eq. \eqref{PPFA}), for the same distances and choices of temperatures of Fig. \eqref{Fig3}.}\label{Fig5}\end{figure}\end{center}

In Fig. \ref{Fig5} we plot the ratio between the exact pressure and the PFA results for the four temperature configurations used in Fig. \ref{Fig3}. We observe that PFA provides in our range of distances a description of the pressure with a relative error typically well below 20\%. The fact the PFA predicts a change of sign not exactly at the position predicted by the exact calculation results in the existence of a vertical asymptote of the ratio $P/P_\text{PFA}$, clearly shown in the blue and orange curves in Fig. \ref{Fig5}.

\subsection{Dependence on geometrical parameters}\label{SecGeoPar}

It is now interesting to understand how a modification of the geometrical parameters of the gratings is able to tune the value of the pressure. To this aim we have chosen as a reference the pressure at a distance $d=4\,\mu$m for $(T_1,T_2,T_\text{e})=(200,400,10)\,$K, for which the pressure is around $P_0=-10^{-6}\,$N\,m$^{-2}$ (see Fig. \ref{Fig3}). Starting from this result, we have modified one by one the values of the filling factor $f$, period $D$, corrugation depth $h$ and calculated the ratio between the modified pressure and the reference $P_0$.

\begin{center}\begin{figure}[htb]
\includegraphics[width=8cm]{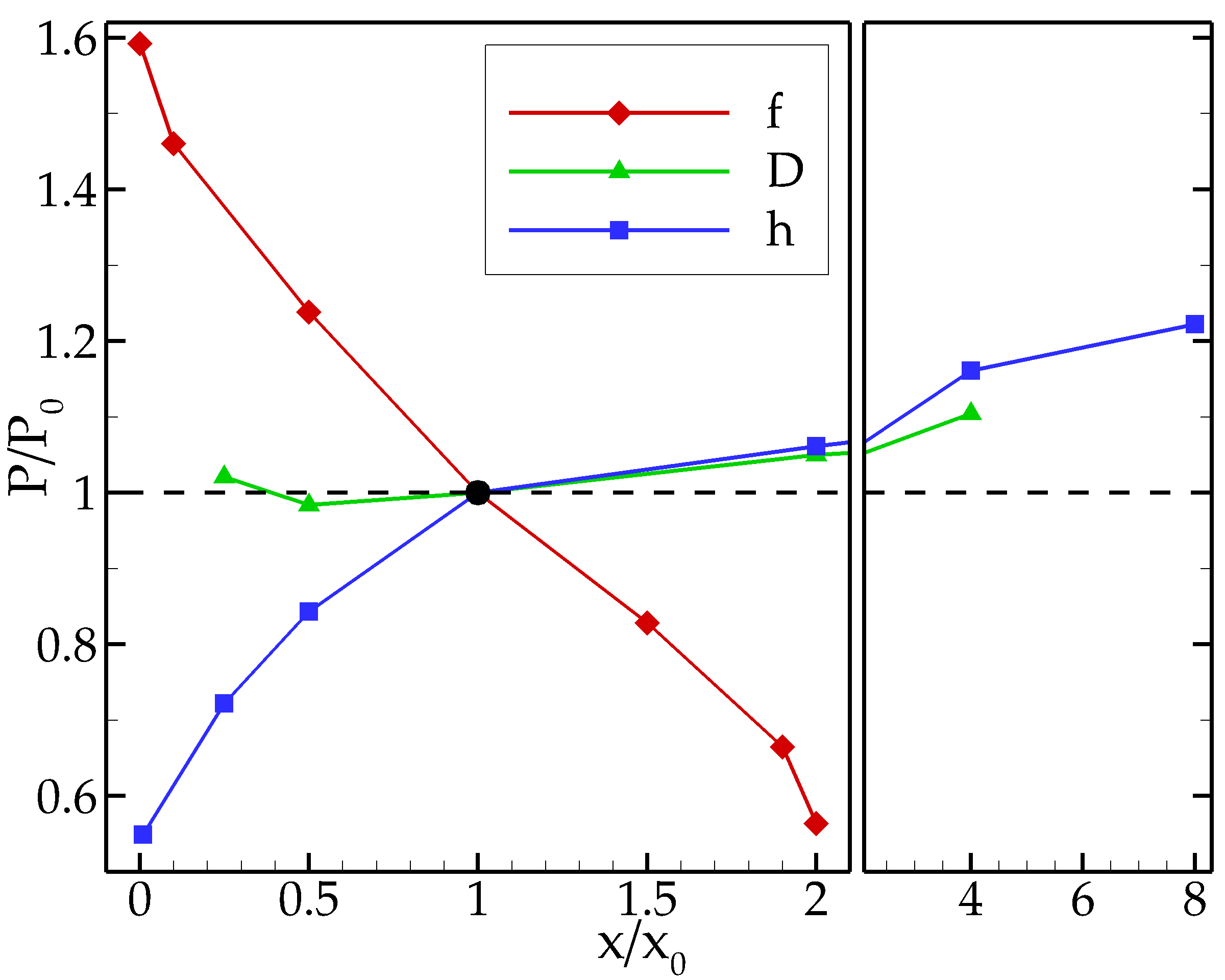}
\caption{(Color online) Variation of the pressure between two gratings at $d=4\,\mu$m [temperatures $(T_1,T_2,T_\text{e})=(200,400,10)\,$K] as a function of the geometrical parameters. The reference point (black circle) corresponds to the set of parameters $f_1=f_2=0.5$, $h_1=h_2=1\,\mu$m, $\delta_1=10\,\mu$m, infinite $\delta_2$, $D=1\,\mu$m. The three curves show the variation of pressure when changing one parameter at a time (red diamonds for the filling factor, green triangles for the period, blue squares for the corrugation depth). On the $y$ axis, the pressures are normalized with respect to the reference one, while on the $x$ axis each varying parameter is normalized with respect to its reference value ($f_0=0.5$, $D_0=1\,\mu$m, $h_0=1\,\mu$m). Note that the plot on the right side continues the one on the left with a modified $x$ scale.}\label{Fig6}\end{figure}\end{center}

\begin{center}\begin{figure}[h!]
\includegraphics[width=8cm]{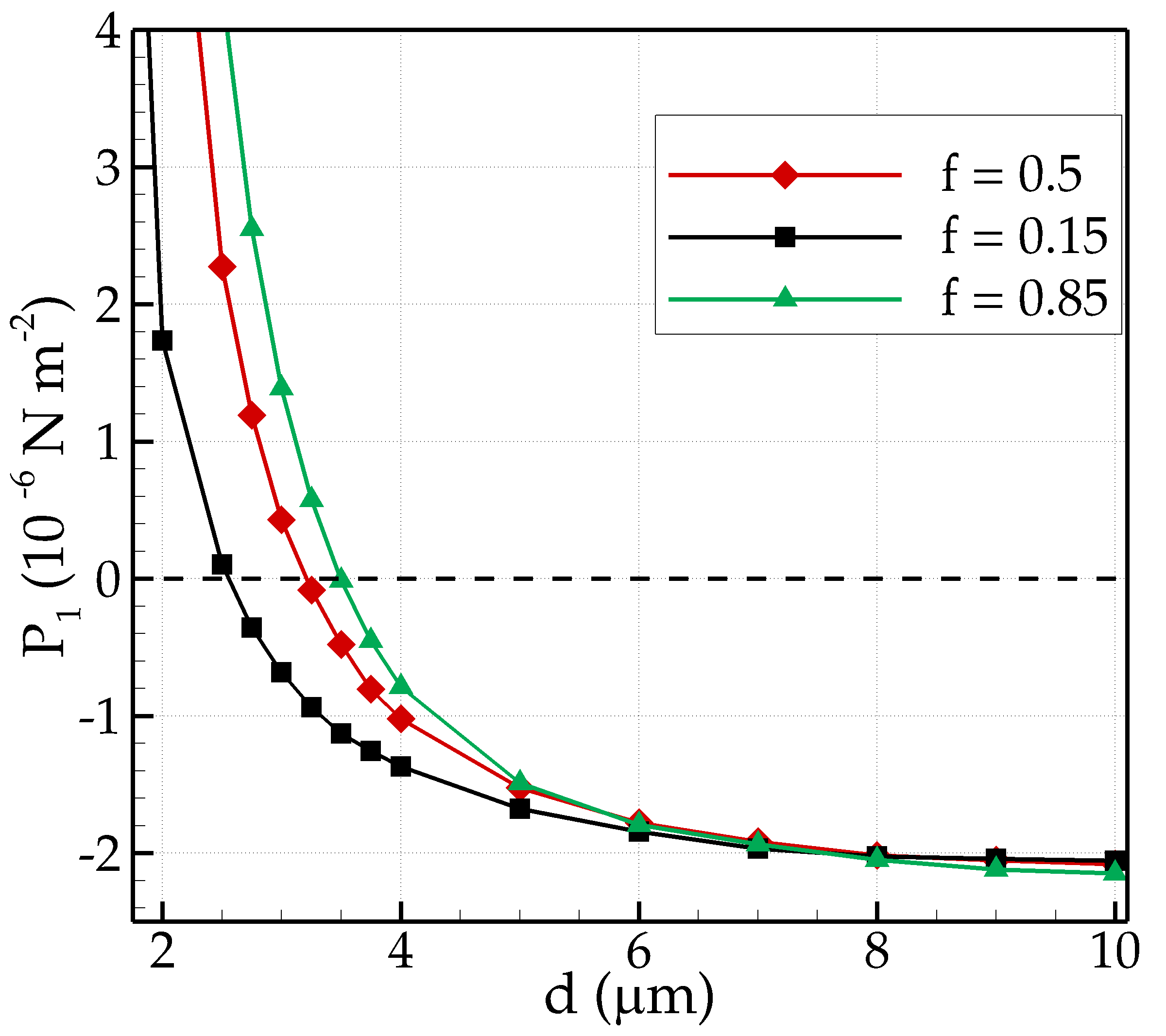}
\caption{(Color online) Pressure on grating 1 as a function of distance [temperatures $(T_1,T_2,T_\text{e})=(200,400,10)\,$K] for three different values of filling factor, all the other geoemtrical parameters being the reference ones.}\label{Fig7}\end{figure}\end{center}

The results are shown in Fig. \ref{Fig6}, where the pressure ratio is plotted as a function of the ratio between the modified parameter and the reference ones ($f_0=0.5$, $D_0=1\,\mu$m and $h_0=1\,\mu$m). First, we observe that geometrical modifications can tune the pressure by a factor going from 0.5 to 1.6. In particular, this region can be fully explored by varying the filling factor between the two admitted extreme values $f=0$ and $f=1$, i.e. between the two limiting slab-slab configurations. Concerning the depth $h$, it also allows a wide variation of the pressure. We remark that for $h$ going to zero we recover the result corresponding to $f=1$, that is a filled grating. On the contrary, for increasing values of $h$, we see that we approach to a pressure approximately equal to half the value of the pressure for $f=1$. This can be interpreted by noticing that roughly speaking at some point the corrugation is so deep that only the upper part (half of the total surface, being $f=0.5$) contributes to the pressure. Differently, the dependence of the pressure on the period $D$ is less pronounced, and absent within our accuracy in the case of a lateral shift between the gratings, not reported in figure.

As we have shown, the filling factor is a promising tool to tailor the behavior of the pressure. This is further pointed out in Fig. \ref{Fig7}, where the distance-dependent pressure is plotted for three different values of $f$. Whereas the asymptotic value of the pressure is practically the same, we note that for small distances the three curves differ visibly. More interestingly, the attractive-repulsive transition can be tuned approximately from 2.5 to 3.5\,$\mu$m by changing $f$ from 0.15 to 0.85.

\subsection{Spectral properties of the pressure}

Let us focus now on the spectral properties of the pressure, by analyzing the quantity $\Delta(\omega)$, defined as the spectral component at frequency $\omega$ of the non-equilibrium contribution to the force \eqref{Delta}, that is
\begin{equation}\label{Deltaomega}\Delta(T_1,T_2,T_\text{e})=\int_0^{+\infty}\hspace{-.2cm}d\omega\,\Delta(\omega).\end{equation}
Also in this case, we consider our reference point $d=4\,\mu$m and $(T_1,T_2,T_\text{e})=(200,400,10)$ and compare its spectral distribution with the two slab-slab cases ($f=0$ and $f=1$) as well as with some variations of one of the three parameters discussed above.

\begin{center}\begin{figure}[htb]
\includegraphics[width=8cm]{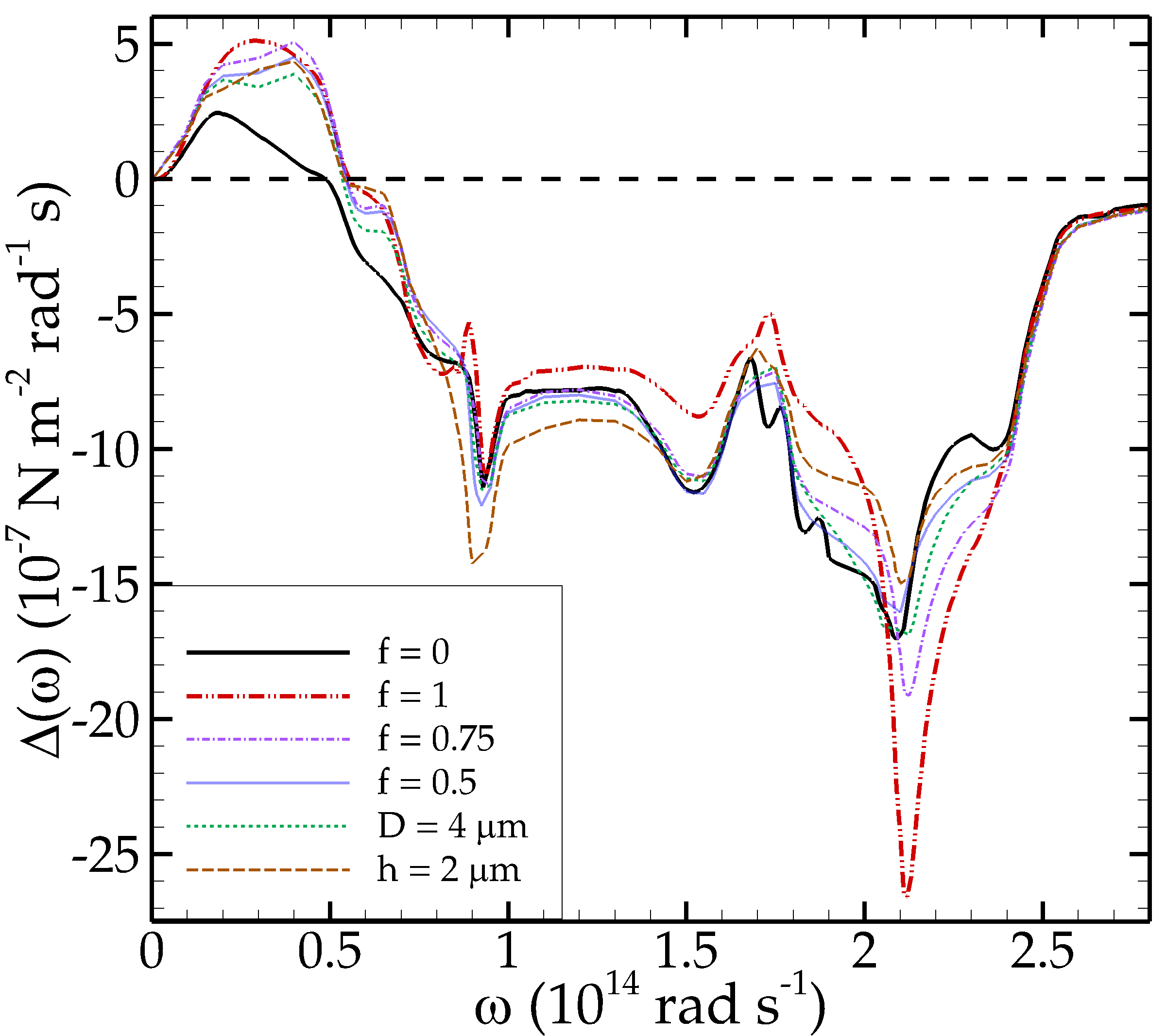}
\caption{(Color online) Spectral density of the OTE contribution to the force (defined in Eq. \eqref{Deltaomega}) at $d=4\,\mu$m [temperatures $(T_1,T_2,T_\text{e})=(200,400,10)\,$K]. The solid black line corresponds to filled gratings ($f=1$), the dot-dot-dashed red line to empty ones ($f=0$), the dotted blue line to our reference gratings, having $f=0.5$. In the other curves we vary the geometrical parameters one by one with respect to our reference case: dot-dashed violet line for $f=0.75$, short-dashed green line for $D=4\,\mu$m, long-dashed brown line for $h=2\,\mu$m.}\label{Fig8}\end{figure}\end{center}

The result is shown in Fig. \ref{Fig8}. We see that no striking spectral difference is present between the configurations compared. Roughly speaking, no new modes (such as the spoof plasmons observed in metal gratings \cite{GarciaVidalJOptA05,YuNatMat10}) are observed in the spectral region of interest, that is up to $\omega$ of the order of $3\times10^{14}\,$rad\,s$^{-1}$. The spectral properties for any considered value of the geometrical parameters show small differences with respect to the ones of the two slab-slab configurations.

\begin{center}\begin{figure}[htb]
\includegraphics[width=8cm]{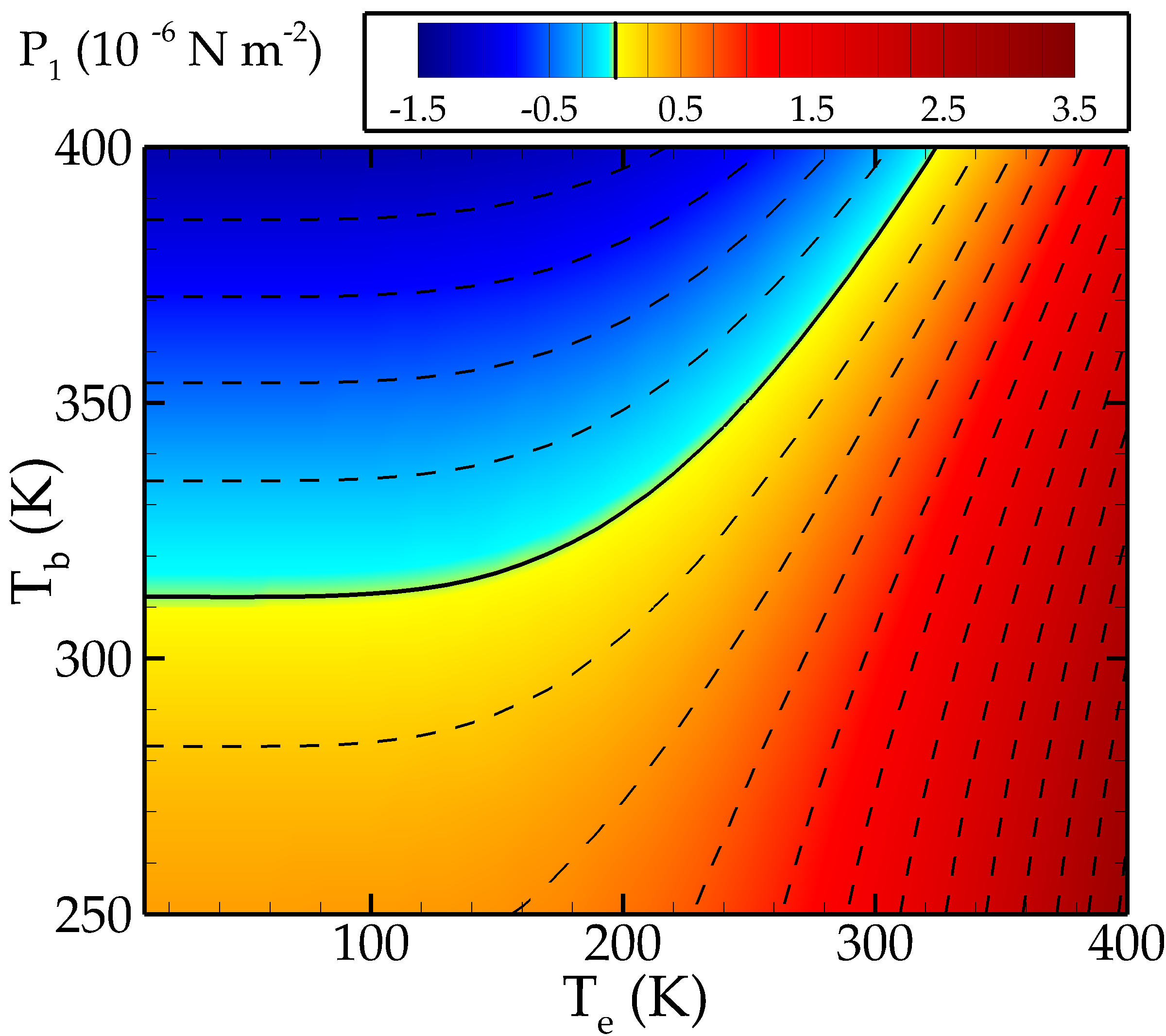}
\caption{(Color online) Pressure acting on grating 1 for $d=4\,\mu$m and $T_1=T_2=T_\text{b}$ as a function of $T_\text{b}$ and $T_\text{e}$. The solid line corresponds to zero pressure, while the dashed lines to the other contour lines shown in legend.}\label{Fig9}\end{figure}\end{center}

\subsection{Modulation of the attractive-repulsive transition}

As we have seen in Sec. \ref{SecGeoPar}, the filling factor is a promising tool to shift the distance at which the transition between attraction and repulsion takes place. Nevertheless, from an experimental point of view it is more interesting to understand how this transition can be affected by tuning parameters which can be varied during an experiment, such as the three temperatures. This is topic of this Section, where we first consider the case in which the two gratings have a common temperature $T_1=T_2=T_\text{b}$, in general different from the environmental one $T_\text{e}$. For this configuration, we plot in Fig. \ref{Fig9} the pressure acting on grating 1 in the reference configuration discussed above as a function of $T_\text{b}$ and $T_\text{e}$.

The plot is clearly divided in two regions, corresponding to positive and negative values of the pressure, separated by a solid zero-pressure line. Following this line, we see that repulsion can be obtained only for body temperatures larger than approximately 312\,K, and that for larger values of $T_\text{b}$ a larger region of $T_\text{e}$ realizes repulsion. Moreover, we stress the remarkable feature that for values of $T_\text{b}$ close to 312\,K the pressure is approximately zero and almost independent on the environmental temperature for $T_\text{e}$ up to approximately 150\,K.

In the same spirit of our last analysis we now fix only $T_1$ at three different values (200, 300 and 400\,K) and let $T_2$ and $T_\text{e}$ vary. The pressure as a function of the two temperatures is shown in Fig. \ref{Fig10}. We see a behavior similar to the one observed in Fig. \ref{Fig9}, that is the existence of a minimum temperature $\bar{T}_2$ below which repulsion is impossible, as well as a region where the pressure is close to zero almost independently of $T_\text{e}$. As manifest from Fig. \ref{Fig10}, the limit temperature $\bar{T}_2$ is a decreasing function of $T_1$.

\begin{widetext}
\begin{center}
\begin{center}\begin{figure}[htb]
\includegraphics[width=18cm]{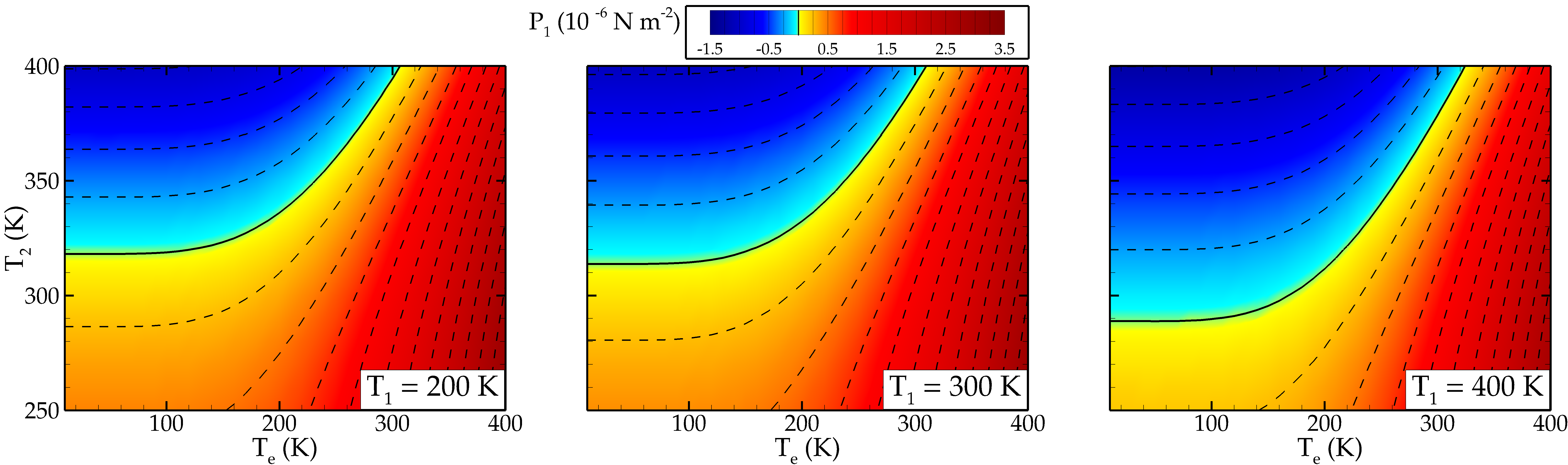}
\caption{(Color online) Pressure acting on grating 1 for $d=4\,\mu$m as a function of $T_2$ and $T_\text{e}$ for three different values of $T_1$. Same convention of Fig. \ref{Fig9} for contour lines.}\label{Fig10}\end{figure}\end{center}
\end{center}
\end{widetext}

\begin{center}\begin{figure}[h!]
\includegraphics[width=8cm]{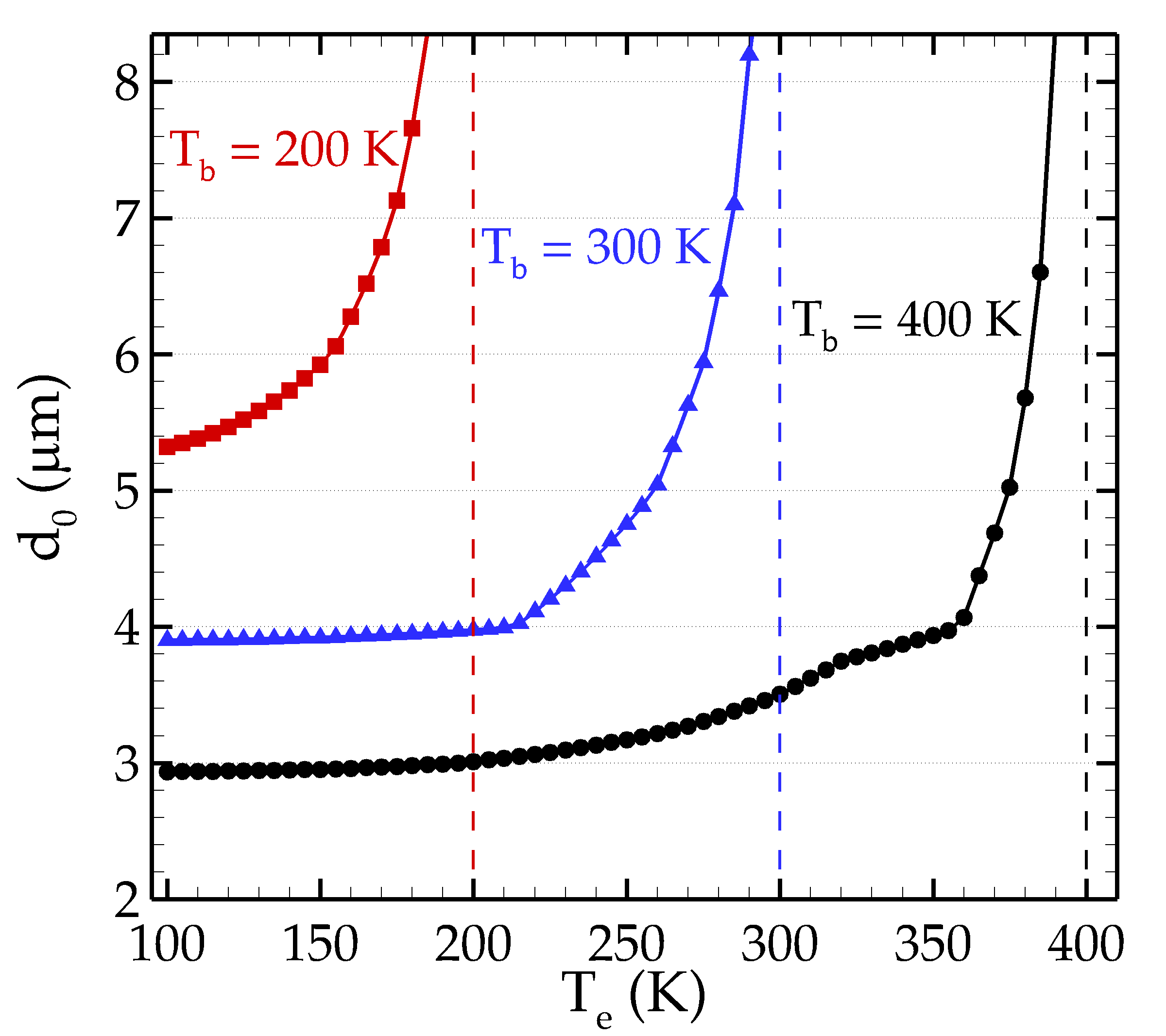}
\caption{(Color online) Distance $d_0$ of attractive-repulsive transition of the pressure as a function of $T_\text{e}$ for three different values of $T_1=T_2=T_\text{b}$.}\label{Fig11}\end{figure}\end{center}

Finally, we discuss how the distance $d_0$ at which the attractive-repulsive transition takes place can be tuned by changing the three temperatures. This is shown in Fig. \ref{Fig11}, where we fix $T_1=T_2=T_\text{b}\in\{200,300,400\}\,$K and plot $d_0$ as a function of $T_\text{e}$. As a general remark, when $T_\text{e}$ is smaller than $T_\text{b}$ the distance $d_0$ tends to a constant value, which decreases from 5.5\,$\mu$m to 3\,$\mu$m for $T_\text{b}$ going from 200 to 400\,K. Furthermore, when $T_\text{e}$ tends to $T_\text{b}$, i.e. the system approaches thermal equilibrium, $d_0$ tends to a vertical asymptote, in accordance to the fact that the pressure is always attractive at thermal equilibrium.

\section{Conclusions}\label{SecConcl}

We calculated the Casimir-Lifsthiz pressure out of thermal equilibrium acting on a 1D dielectric lamellar grating in front of another (in general different) dielectric grating. To this aim, we implemented the Fourier Modal Method in order to derive the scattering operators associated to each individual grating. Using the general formalism for Casimir-Lifshitz force based on scattering matrices, we calculated the pressure acting on a finite Fused Silica grating in presence of an infinite Silicon grating, and also compared our results to the Proximity Force Approximation.

We showed that the combination of geometrical structuring of the surface and absence of thermal equilibrium offers an extremely rich domain of variation both with respect to thermal equilibrium and with respect to planar slabs out of thermal equilibrium. As in the case of two slabs, non-equilibrium is able to produce a repulsive pressure, whose intensity can be tuned by varying the temperatures as well as the several geometrical parameters associated to each grating. We also pointed out the presence of regimes in which the pressure is close to zero and almost independent of the environmental temperature. Remarkably, the variations of all the parameters strongly affect the distance at which the transition between attractive and repulsive pressure occurs, allowing to obtain transition distances as low as $2.5\,\mu$m. This feature is indeed promising for the experimental observation of a repulsive force. Moreover, our results can be relevant in the context of force manipulations on micro-mechanical systems \cite{ZouNatComm13}. Finally, an extension of this study to three-body configurations is also promising toward the manipulation of heat transfer \cite{MessinaPRL12,MessinaPRA14}.

\begin{acknowledgments}
The authors acknowledge financial support from the Julian Schwinger Foundation.
\end{acknowledgments}

\end{document}